\documentclass[twocolumn, tighten, twocolappendix]{aastex631_mod}
\usepackage[T1]{fontenc}
\usepackage{microtype}
\usepackage{newtxtext}
\usepackage[varvw]{newtxmath}
\usepackage{mathtools}
\usepackage{enumitem}
\usepackage{booktabs}
\usepackage{tabularx}
\usepackage[para,referable]{threeparttablex}
\usepackage{multirow}
\usepackage{tikz}
\usepackage{bm}

\usetikzlibrary{shapes}
\hypersetup{pdfauthor={Errani et al.},
            pdftitle={},
            pdfkeywords={},
            bookmarksnumbered=true}

\newcommand{\Trel}{T_\mathrm{rel}}
\newcommand{\kms}{\mathrm{km\,s^{-1}}}
\newcommand{\Rh}{R_\mathrm{h}}

\newcommand{\rh}{r_\mathrm{h}}
\newcommand{\Mh}{M_\mathrm{h}}

\newcommand{\diff}{\mathrm{d}}
\newcommand{\Msol}{\mathrm{M_{\odot}}}

\newcommand{\kpc}{\mathrm{kpc}}
\newcommand{\pc}{\mathrm{pc}}
\newcommand{\Gyr}{\mathrm{Gyr}}
\newcommand{\Myr}{\mathrm{Myr}}

\newcommand{\g}{\mathrm{g}}
\newcommand{\E}{\mathcal{E}}
\newcommand{\plus}[1] {^{\mathmakebox[\widthof{$^-$}][c]{+}#1}}
\newcommand{\minus}[1]{_{\mathmakebox[\widthof{$^-$}][c]{-}#1}}

\graphicspath{{./}{./Figures/}}

\begin{document}
\shorttitle{Stellar Mass Segregation in Dark Matter Halos}
\shortauthors{Errani, Pe\~narrubia \& Walker}
\title{Stellar Mass Segregation in Dark Matter Halos}

\author{Rapha\"el Errani}
\affiliation{McWilliams Center for Cosmology and Astrophysics, Department of Physics, Carnegie Mellon University, Pittsburgh, PA 15213, USA}
\email{errani@cmu.edu}

\author{Jorge Pe\~narrubia}
\affiliation{Institute for Astronomy, University of Edinburgh, Royal Observatory, Blackford Hill, Edinburgh EH9 3HJ, UK}

\author{Matthew G. Walker}
\affiliation{McWilliams Center for Cosmology and Astrophysics, Department of Physics, Carnegie Mellon University, Pittsburgh, PA 15213, USA}

\received{May 28 2025}
\revised{August 25 2025}
\accepted{September 2 2025}

\begin{abstract}
We study the effect of stellar mass segregation driven by collisional relaxation within the potential well of a smooth dark matter halo. This effect is of particular relevance for old stellar systems with short crossing times, where small collisional perturbations accumulate over many dynamical timescales. We run collisional $N$-body simulations tailored to the ambiguous stellar systems Ursa Major~3/Unions~1, Delve~1 and Eridanus~3, modeling their stellar populations as two-component systems of high- and low-mass stars, respectively. For Ursa Major~3/Unions~1 (Delve~1), assuming a dynamical-to-stellar mass ratio of 10, we find that after 10 Gyr of evolution, the radial extent of its low-mass stars will be twice as large as (40 per cent larger than) that of its high-mass stars. We show that weak tides do not alter this relative separation of half-light radii, whereas for the case of strong tidal fields, mass segregation facilitates the tidal stripping of low-mass stars. We further find that as the population of high-mass stars contracts and cools, the number of dynamically formed binaries within that population increases. Our results call for caution when using stellar mass segregation as a criterion to separate star clusters from dwarf galaxies, and suggest that mass segregation increases the abundance of massive binaries in the central regions of dark matter-dominated dwarf galaxies.
\end{abstract}

\section{Introduction}
\label{Sec:Intro}
In cold dark matter cosmology, galaxies are expected to form deep within the potential wells of dark matter halos \citep{WhiteRees1978}. Numerical simulations suggest that these cold dark matter halos reach remarkably high central densities, well-described by a universal centrally-divergent density profile \citep{Navarro1996a, Navarro1997}. The centrally-divergent centers of cold dark matter halos are commonly referred to as ``cusps''. Density cusps render cold dark matter halos resilient to the effect of tides \citep{Penarrubia2010, vdb2018}: for a fixed tidal field strength, it is argued that cold dark matter halos cannot be tidally stripped beyond a certain point and instead converge toward a stable asymptotic remnant state \citep{EN21, Stuecker2023}. 
Stars embedded in such cold dark matter cusps would be protected from tidal disruption, plausibly giving rise to a population of ``micro galaxies'' \citep{EP20, EINPW2024}. 
The discovery of such objects would allow us to put strong constraints on the nature of dark matter, as discussed in the context of a potential self-annihilation signal \citep{Crnogorcevic2024, ENSM24}, primordial black hole dark matter \citep{Graham2024} or ultra-light particle dark matter \citep{Safarzadeh2020}.

In recent years, deep photometric surveys have led to the discovery of several objects with structural parameters at the interface of the globular cluster and dwarf galaxy regimes \citep{Conn2018,Mau2020,Cerny22,Cerny23,Smith2024}. Could some of these systems possibly be among the faintest dark matter-dominated dwarf galaxies known to date \citep{ENSM24,Smith2024,Simon2024}?
Proving unequivocally that these objects are either dark matter-dominated dwarf galaxies or star clusters devoid of any dark matter has turned out to be a very challenging task. 

Traditionally, stellar kinematics have been used to argue in favor of the high dark matter content of dwarf galaxies \citep{Mateo1998, Walker2007}. For faint stellar systems with few member stars, the inferred velocity dispersions are shown to depend sensitively on the inclusion or exclusion of individual members stars \citep{Smith2024} and the choice of prior \citep{Simon2024}. The presence of binary stars further complicates such studies by adding a velocity dispersion floor that, particularly for low-mass dwarf galaxies, needs to be accurately accounted for and generally requires the availability of multi-epoch spectroscopic measurements \citep{McConnachieCote2010, Buttry2022}. 

Another pathway suggested to distinguish dwarf galaxies from globular clusters has been to use the elemental abundance patterns of their stars (see, e.g., \citealt{Gratton2012}, \citealt{BastianLardo2018} for a discussion of elemental abundances in globular clusters, and \citealt{Venn2004}, \citealt{Ji2019_Gru1} for dwarf galaxies), though their application to faint stellar systems with few member stars remains challenging (see, e.g., \citealt{Zaremba2025}; see also \citealt{Fu2023} who use narrow-band imaging to infer metallicity dispersions for faint stellar systems where spectroscopic measurements are not feasible).

A further strategy discussed in the literature is based on tidal survival: the existence of the ancient stellar system Ursa Major~3/Unions~1 \citep{Smith2024} in the inner region of the Milky Way has been used to argue in favour of it being embedded in a dark matter halo and in turn protected from tidal disruption \citep{ENSM24}. This picture, though, has recently been challenged by \citet{Devlin2025}, who argue that the baryonic mass of Ursa Major~3/Unions~1 has been underestimated in its discovery paper, making it more stable against Milky Way tides even in the absence of a surrounding dark matter halo. 

Yet another method to distinguish dark matter-dominated objects from those without dark matter has been to search for observational signatures of collisional and collisionless dynamics. 
The canonical picture is that in dark matter-dominated systems, stellar orbits are determined by the (dark matter) mean field and obey the collisionless Boltzmann equation. Hence, particle-mesh \citep{Fellhauer2000} and tree codes with force softening \citep{Springel2005Gadget} have been employed for their study. For globular clusters instead, the importance of close stellar encounters is thought to play an important role in shaping their complex dynamical evolution \citep{Spitzer1987book}, with direct $N$-body codes being necessary for their study \citep{Aarseth1999}. 
A prominent signature of collisional processes is the segregation of stellar masses, which has been discussed also in the context of the nature of faint stellar systems \citep{Kim2015, Baumgardt2022, Simon2024, Zaremba2025} and as a potential means to constrain the progenitors of tidal streams with seemingly conflicting dynamical and chemical signatures\footnote{The C-19 stellar stream \citep{Martin2022_C19} has width of ${\sim}160\,\pc$ and a velocity dispersion of ${\sim}6\,\kms$ \citep{Yuan2022}, hinting at a dwarf galaxy origin \citep{ENIM2022}. This appears to be in conflict with the near-zero metallicity spread of its member stars as well as anti-correlations in its elemental abundances, typically seen in globular clusters.} \citep{ENIM2022}.

In this work, we challenge the canonical picture that dark matter-dominated systems do not show signatures of collisional processes. Taking the ambiguous stellar systems 
Ursa Major~3/ Unions~1 (\texttt{UMa3/U1} for short, stellar mass $M_\star = 16\plus{6}\minus{5}\,\Msol$, projected half-light radius $\Rh=(3\pm1)\pc$, line-of-sight velocity dispersion $\sigma_\mathrm{los} \lesssim 4\,\kms$, see \citealt{Smith2024} and footnote~\ref{Footnote:UMa3}%
\footnotetext{For \texttt{UMa3/U1}, \citet{Smith2024} find $\sigma_\mathrm{los} = 3.7\plus{1.4}\minus{1.0}\,\kms$ when including all member stars in their analysis, while the dispersion drops to $1.9\plus{1.4}\minus{1.1}\,\kms$ when excluding the furthest outlier, and is unresolved when excluding one additional star, see their Fig.~5.\label{Footnote:UMa3}}%
), Delve\,1 (\texttt{Del1}, $M_\star = 144\plus{24}\minus{27}\,\Msol $, $\Rh =6.2\plus{1.5}\minus{1.1}\,\pc $, $\sigma_\mathrm{los}\lesssim 1.2\,\kms$ see \citealt{Mau2020}, \citealt{Simon2024} and footnote~\ref{Footnote:Del1Eri1}%
\footnotetext{The velocity dispersions for \texttt{Del1} and \texttt{Eri3} listed here are the 90 per cent confidence upper limits of \citealt{Simon2024} inferred using log-uniform priors. For (linearly) uniform priors, the upper limits are $\sigma_\mathrm{los} \lesssim 2.5\,\kms$ and $\lesssim 9.1\,\kms$, respectively.\label{Footnote:Del1Eri1}}%
) and Eridanus\,3 (\texttt{Eri3}, $M_\star = 800\plus{470}\minus{300}\,\Msol$, $\Rh=8.6\plus{0.9}\minus{0.8}\,\pc$, $\sigma_\mathrm{los}\lesssim 5.4\,\kms$, see \citealt{Conn2018}, \citealt{Simon2024} and footnotes~\ref{Footnote:Del1Eri1} and \ref{Footnote:Eri1}) \footnotetext{We estimate the stellar mass of \texttt{Eri3} from its published luminosity, assuming a stellar mass-to-light ratio of $\Upsilon_\star=1.4$. This value is chosen to match the stellar mass-to-light ratios of \texttt{UMa3/U1} \citep{Smith2024} and \texttt{Del1} \citep{Mau2020}.\label{Footnote:Eri1}} as examples, we run collisional $N$-body simulations where we assume that these systems are dark matter-dominated and deeply embedded in a smooth dark matter halo. We will show that, driven by their short crossing times, small collisional perturbations due to the minute potential fluctuations caused by their own stars' gravity can sum up over many gigayears and influence their dynamical evolution. Specifically, for \texttt{UMa3/U1} and \texttt{Del1}, we show that signatures of mass segregation can be observed even for dynamical-to-stellar mass ratios as high as ${\sim} 50$ and ${\sim} 20$, respectively.

The paper is structured as follows. In section~\ref{Sec:RelaxationTimes}, we estimate the timescale for collisional relaxation in the presence of a smooth and static dark matter halo. In section~\ref{sec:NumericalSetup}, we detail the numerical setup of our collisional $N$-body simulations. In section~\ref{Sec:Static}, we discuss the results of our simulations for the case of a static dark matter halo surrounding each stellar system, and in Sec.~\ref{sec:Binaries} we describe the dynamical formation of stellar binaries. We extend our analysis to a broader range of initial dynamical-to-stellar mass ratios in Sec.~\ref{Sec:DynamicalToStellarRatio}. In Sec.~\ref{Sec:Tides}, we study the effect of galactic tides on stellar mass segregation through a time-evolving dark matter potential. Finally, we summarize our results and conclusions in section~\ref{sec:conclusions}.

\section{Relaxation Times}
\label{Sec:RelaxationTimes}

The faint stellar systems \texttt{UMa3/U1}, \texttt{Del1} and \texttt{Eri3} host ancient stellar populations, with isochrone fits suggesting stellar ages beyond $10\,\Gyr$. For such old systems, it seems plausible that small dynamical perturbations may accumulate over time and give rise to some secular evolution. As we will show in the following, this idea holds even if the stellar population is embedded in a dark matter potential. 

If the stellar contribution to the total potential is fully negligible and the dark matter potential is smooth, then the system obeys the Collisionless Boltzmann Equation and, in absence of other perturbations, no secular evolution occurs. However, one may image a system where the stellar contribution to the potential becomes relevant for its dynamical evolution by providing a noisy, fluctuating component to the potential. To illustrate the relevant timescales at play, we will now estimate the timescale for collisional relaxation due to gravitational encounters between stars in presence of a smooth dark matter potential. 

We call $\Mh = M_\mathrm{sub}({<}\rh) + M_\star({<}\rh) $ the total dynamical mass enclosed within the (3D) stellar half-light radius $\rh$, which is the sum of the dark matter subhalo mass $M_\mathrm{sub}({<}\rh)$ and the stellar mass $M_\star({<}\rh)=M_\star/2$. For short, we will refer to the average dynamical-to-stellar mass ratio within $\rh$ as 
\begin{equation}
 \Upsilon_\mathrm{dyn} \equiv \Mh / M_\star({<}\rh)~.
\end{equation}
The dynamical-to-stellar mass ratio $\Upsilon_\mathrm{dyn}$ can be observationally constrained via the line-of-sight velocity dispersion $\sigma_\mathrm{los}$ using the relation $\Upsilon_\mathrm{dyn} \approx 8 \Rh \sigma^2_\mathrm{los} G^{-1} M_\star^{-1}$, where $\Rh$ is the projected half-light radius (see Eq.~8 of \citealt{ENSM24}). For the systems \texttt{UMa3/U1}, \texttt{Del1}, and \texttt{Eri3}, the upper limits on their respective velocity dispersions (as listed in Sec.~\ref{Sec:Intro}) yield approximate constraints of $\Upsilon_\mathrm{dyn} \lesssim 5600$, $120$ and $580$, respectively. If these objects are indeed dark matter-dominated relics of the Milky Way's early accretion history, their true dynamical-to-stellar mass ratios could be significantly lower than the current upper bounds. Taking \texttt{UMa3/U1} as an example, its pericentre and apocentre of ${\sim}13\,\kpc$ and ${\sim}26\,\kpc$ \citep{Smith2024}, respectively, suggest an early accretion onto the Milky Way (see Fig.~1 of \citealt{EINPW2024}). This makes it plausible that the subhalo surrounding \texttt{UMa3/U1} is structurally close to the maximally stripped ``tidal remnant'' state for its orbit \citep{EN21}, corresponding to a mean enclosed dark matter density of order $\bar \rho_\mathrm{h} \sim 10^9\,\Msol\,\kpc^{-3}$ and a dynamical-to-stellar mass ratio of order $\Upsilon_\mathrm{dyn} \sim 30$ (see Fig.~4 and 5 of \citealt{ENSM24}). As a working example, we adopt a fiducial\footnote{Note that the total mass-to-luminosity ratio $M_\mathrm{tot}/L$, expressed in units of $\Msol/\mathrm{L}_\odot$, may be substantially higher than the value of $\Upsilon_\mathrm{dyn}$, and depends sensitively on how much the dark matter subhalo extends beyond the stellar half-light radius.} value of $\Upsilon_\mathrm{dyn} = 10$ and later explore a broader range, $3 \lesssim \Upsilon_\mathrm{dyn} \lesssim 300$, in Sec.~\ref{Sec:DynamicalToStellarRatio}.

A stellar population that is embedded in a dark matter subhalo has a (3D) velocity dispersion of roughly $ \langle v^2 \rangle \approx  G \Mh / \rh  = G M_\star \Upsilon_\mathrm{dyn} / (2\rh) $,
and a crossing time of
\begin{equation}
 T_\mathrm{cross} = \rh^{3/2} \left( G \Mh \right)^{-1/2} =  \sqrt{2/G}~\rh^{3/2} \left(M_\star  \Upsilon_\mathrm{dyn} \right)^{-1/2} ~. \label{eq:Tcross}
\end{equation}
For the case of UMa3/U1, for example, we find $T_\mathrm{cross} \approx 13\,\Myr$ when assuming a dynamical-to-stellar mass ratio $\Upsilon_\mathrm{dyn}=10$: over $10\,\Gyr$, the system goes through close to ${\sim} 800$ crossing times. Over one crossing time, the average increase in squared velocity that an individual star experiences due to the fluctuating gravitational potential of $N_\star$ stars with individual masses $m_\star$ can be approximated by (using Eq.~18 of \citealt{Penarrubia2019Scattering} and assuming that the stellar number density $\bar n=3N_\star/(8\pi \rh^3)$ is approximately constant within the half-light radius)
\begin{eqnarray}
 \langle \Delta v^2 \rangle_{t=T_\mathrm{cross}} / \langle v^2 \rangle &\approx&  \sqrt{24\pi}~ \Upsilon_\mathrm{dyn}^{-2} N_\star^{-1}   \left[ \ln(\Lambda) - 1.9\right] \label{Eq:DeltaV2}\\
                                                  &=& \sqrt{3\pi/2} ~ \Mh^{-2} N_\star m_\star^2  \left[ \ln(\Lambda) - 1.9\right]  \label{Eq:DeltaV2_Nm2}
 \label{eq:v2_kick}
 \end{eqnarray}
where $v \approx \langle v^2 \rangle^{1/2}$ is the velocity of a star, and $\ln(\Lambda)$ is the Coulomb logarithm\footnote{We adopt a constant $\ln(\Lambda) = 8.2$ as suggested by the simulation results of \citet[][see their Fig.~3]{Penarrubia2019Scattering}.}. From Eq.~\ref{Eq:DeltaV2} wee see that, all else equal, $\langle \Delta v^2 \rangle$ decreases as $\Upsilon_\mathrm{dyn}$ and $N_\star$ increase: for $\Upsilon_\mathrm{dyn} \rightarrow \infty$ and $N_\star \rightarrow \infty$, the system becomes collisionless.

As time progresses, these $\langle \Delta v^2 \rangle$ accumulate. The orbital motion of a star is driven by the potential fluctuations once $\langle \Delta v^2 \rangle / \langle v^2 \rangle \approx 1$, which happens over the course of a relaxation time
\begin{eqnarray}
  T_\mathrm{rel} &\approx&\! \bigl(1/\sqrt{12\pi G}\bigr)\left( \Upsilon_\mathrm{dyn} \rh \right)^{3/2} N_\star M_\star^{-1/2}  \left[ \ln(\Lambda) - 1.9\right]^{-1} \label{Eq:TrelaxUps} \\
                 &=& \sqrt{2/(3\pi G)}\, \left( \Mh \rh \right)^{3/2}  N_\star^{-1} m_\star^{-2} \left[ \ln(\Lambda) - 1.9\right]^{-1}. \label{Eq:TrelaxMhrh}
\end{eqnarray}
All else being equal, the more dark matter-dominated the system is, the longer is its relaxation time. 

For a population of stars with a mass function  $\diff N_\star / \diff m_\star$, the relaxation time will be driven by those stars that maximize the change in $\langle \Delta v^2 \rangle$: Eq.~\ref{Eq:DeltaV2_Nm2} shows that contribution peaks for stars of a stellar mass $m_\star$ that maximizes the product $m_\star^2 \diff N_\star / \diff m_\star$. For a \citet{Chabrier2003} present-day\footnote{We here adopt a \citet{Chabrier2003} present-day mass function as a conservative approximation for the mass function of an older stellar system: this choice of mass function results in a larger relaxation time by not including the collisional effects of massive but short-lived stars.} mass function, stars with masses $0.4 \lesssim m_\star / \Msol \lesssim 1.4$ contribute $68$ per cent of the total $\langle\Delta v^2 \rangle$. More massive stars do not play much of a role by virtue of their low abundance, while less massive stars do not contribute much by virtue of their mass.

\begin{figure}[tb]
 \centering
 \includegraphics[width=8.5cm]{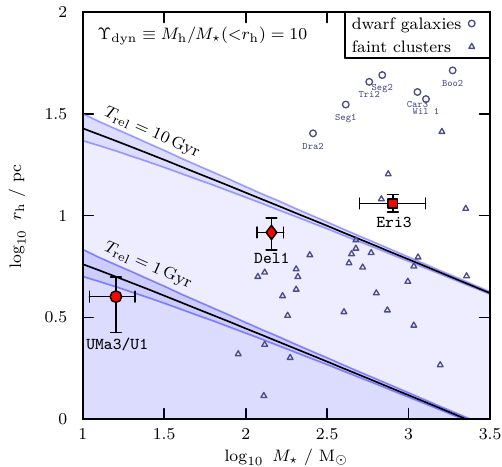}
 \caption{Several Milky Way satellites have relaxation times $\Trel$ shorter than $10\,\Gyr$ even in the presence of substantial amounts of dark matter. Those objects may exhibit stellar mass segregation driven by collisional relaxation. Shown are stellar masses $M_\star$ and (3D) half-light radii $\rh$ of dwarf galaxies (open circles), ambiguous stellar clusters (triangles), as well as the ``micro galaxy'' candidates \texttt{UMa3/U1}, \texttt{Del1} and \texttt{Eri3} (red filled circle, diamond and square, respectively). See footnote~\ref{Footnote:References} for references. Curves of constant relaxation time $T_\mathrm{rel} = 1\,\Gyr$ and $10\,\Gyr$ are computed assuming a dynamical-to-stellar mass ratio of $\Upsilon_\mathrm{dyn} = 10$ within the half-light radius, with stellar masses sampled from a \citet{Chabrier2003} present-day mass function (see text for details).}
 \label{fig:Trelax}
\end{figure}

In Fig.~\ref{fig:Trelax}, we show relaxation times computed by summing Eq.~\ref{eq:v2_kick} over individual stellar masses $m_\star$ drawn from a \citet{Chabrier2003} present-day mass function (PDMF), together with a compilation\footnote{Properties of faint stellar clusters are as compiled in \citet{Cerny22, Cerny23}. The dwarf galaxy data is taken from \citet{McConnachie2012}, version January 2021, with updates for  Bo\"otes~2 \citep{Bruce2023}, Draco~2 \citep{Martin2016_Dra2, Longeard2018} and TriII \citep{Martin2016_Tri2,Kirby2017}. For the faint stellar systems \texttt{UMa3/U1} \citep{Smith2024}, \texttt{Del1} \citep{Mau2020} and \texttt{Eri3} \citep{Conn2018} see Table~\ref{Tab:SimulationParameters} and footnotes~\ref{Footnote:UMa3}, \ref{Footnote:Del1Eri1}, \ref{Footnote:Eri1}.\label{Footnote:References}} of stellar masses $M_\star$ and (3D) half-light radii $\rh$ of Milky Way dwarf galaxies and faint clusters with structural properties at the interface of the globular cluster and dwarf galaxy regimes. In Fig.~\ref{fig:Trelax}, we assume $\Upsilon_\mathrm{dyn}=10$, but the results are easily translated to other choices for dynamical-to-stellar mass ratio through Eq.~\ref{Eq:TrelaxUps}. Uncertainties on the relaxation times shown here stem from sampling the stellar mass function and span the $16^\mathrm{th}$ to $84^\mathrm{th}$ percentiles of crossing times for random realizations of total mass $M_\star$. 

For the faint stellar systems UMa3/U1 and Delve1, assuming a dynamical-to-stellar mass ratio of $\Upsilon_\mathrm{dyn} = 10$, we find relaxation times of $0.9$ and $8.4\,\Gyr$, respectively, substantially shorter than the age of their stellar populations. For these systems, we can therefore expect dynamical signatures of collisional processes, such as stellar mass segregation, even when embedded in a smooth, static and gravitationally dominant dark matter subhalo.

\section{Numerical Setup}
\label{sec:NumericalSetup}
To study the observable effects of collisional relaxation on a stellar population embedded in a smooth dark matter subhalo, we perform a series of $N$-body experiments. In the following, we summarize the details of our numerical setup.

\subsection{Example systems}
We build our $N$-body models to approximately resemble the faint stellar systems \texttt{UMa3/U1} \citep{Smith2024}, \texttt{Del1} \citep{Mau2020} and \texttt{Eri3} \citep{Conn2018}.
These systems are plausible candidates for being the smallest dwarf galaxies known to date, motivated by their kinematics and chemistry \citep{Simon2024}, or by their tidal survival on a low-pericentre orbit \citep{ENSM24}. Table~\ref{Tab:SimulationParameters} lists the structural parameters used for our $N$-body models. We estimate the (3D) half-light radii $\rh \approx 4 \Rh /3$ from the published projected radii $\Rh$. For \texttt{UMa3/U1} and \texttt{Del1}, we use total stellar masses as published; for \texttt{Eri3}, we estimate the stellar mass from its luminosity, assuming a stellar mass-to-light ratio of $\Upsilon_\star=1.4$. This value matches the stellar mass-to-light ratio inferred by \citet{Smith2024} and \citet{Mau2020} for \texttt{UMa3/U1} and \texttt{Del1}, respectively.

\begin{table}[tb]
\begin{threeparttable}
\centering
\caption{Parameters used for the $N$-body models. The table lists the total stellar mass $M_\star$ (for the case of \texttt{Eri3} estimated from its luminosity), the (3D) half-light radius $\rh\approx 4 \Rh/3$ estimated from the projected $\Rh$, and the total number of star particles $N_\star$. For reference, we also list the resulting crossing (Eq.~\ref{eq:Tcross}) and relaxation times (Eq.~\ref{Eq:TrelaxUps}) assuming $\Upsilon_\mathrm{dyn}=10$ and $20$. }
\label{Tab:SimulationParameters}
\begin{tabularx}{\linewidth}{lccc@{\hspace{0.6cm}}cc}
\toprule
\multirow{2}{*}{Model} & \multirow{2}{*}{$M_\star/\Msol$}           & \multirow{2}{*}{$\rh/\pc$}                    &\multirow{2}{*}{$N_\star$} &\multicolumn{2}{c}{for $\Upsilon_\mathrm{dyn} = 10$ $(20)$}\\
                       &                                            &                                               &                           &$T_\mathrm{cross}/\Myr$&$T_\mathrm{rel}/\Gyr$\\ \midrule
\texttt{UMa3/U1}       & 16\tnotex{tn:1}                            & 4\tnotex{tn:1}                                & 50                        &13 (9)                    &0.9 (2.7)                 \\
\texttt{Del1}          & 144\tnotex{tn:2}                           & 8.3\tnotex{tn:2}{}$^\text{,}$\tnotex{tn:3}    & 450                       &13 (9)                    &8.4 (24)                 \\
\texttt{Eri3}          & 800\tnotex{tn:4}$^\text{,}$\tnotex{tn:5}   & 11.5\tnotex{tn:4}                             & 2500                      &9 (6)                     &32 (91)                \\
\bottomrule
\end{tabularx}
\begin{tablenotes}
 \item[a] \label{tn:1}\citet{Smith2024}
 \item[b] \label{tn:2}\citet{Mau2020}
 \item[c] \label{tn:3}\citet{Simon2024}
 \item[d] \label{tn:4}\citet{Conn2018}
 \item[e] \label{tn:5}Assuming $\Upsilon_\star=1.4$, see footnote~\ref{Footnote:Eri1}
\end{tablenotes}
\end{threeparttable}
\end{table}

\begin{figure*}
\centering
\includegraphics[width=18cm]{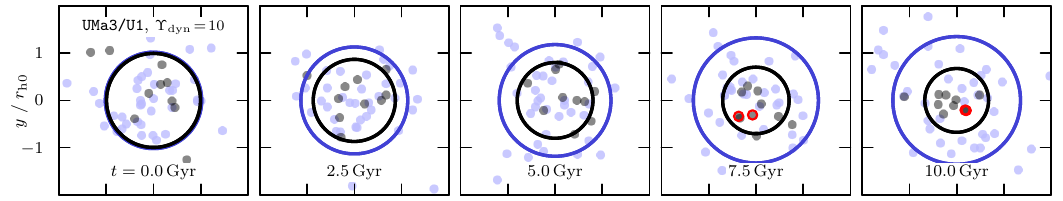}
 \includegraphics[width=18cm]{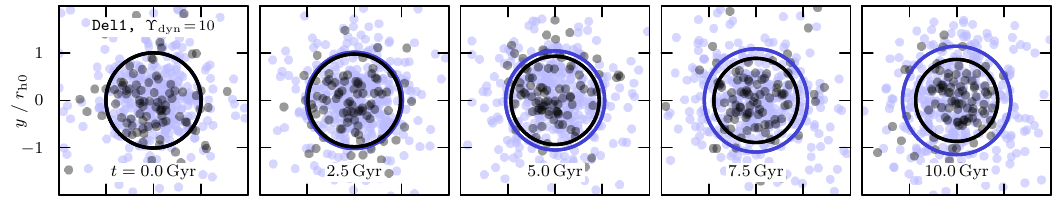}
\includegraphics[width=18cm]{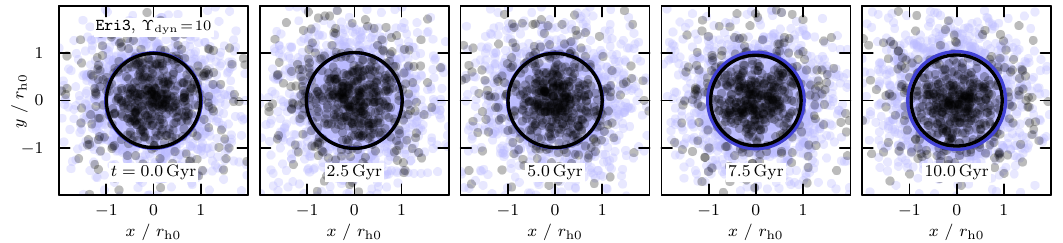}
\caption{Stellar mass segregation in $N$-body realizations of the example systems \texttt{UMa3/U1} (top), \texttt{Del1} (center) and \texttt{Eri1} (bottom). Each system is embedded in a cuspy dark matter halo, with an initial dynamical-to-stellar mass ratio of $\Upsilon_\mathrm{dyn} = 10$. Shown are $\{x,y\}$ projections of snapshots at $t/\Gyr=0$, $2.5$, $5$, $7.5$ and $10$. The axes are expressed in units of the initial half-light radius $r_\mathrm{h0}$. Individual high-mass stars are shown as dark grey points, while low-mass stars are shown in blue. The median half-light radii of high-mass and low-mass stars (computed over a sample of $N$-body realizations, see text) are shown as black and blue circles, respectively. Over a period of $10\,\Gyr$, the half-light radius of the population of low-mass stars expands, while the half-light radius of the high-mass population contracts. Consistent with the relaxation time estimates of Fig.~\ref{fig:Trelax}, the effect is largest for the \texttt{UMa3/U1} model, and smallest for the \texttt{Eri3} model. Dynamically formed binaries consisting of two high-mass stars are highlighted in red (see Sec.~\ref{sec:Binaries}). A video supplementing this figure is available in Appendix~\ref{Appendix:Animations}.}
\label{fig:Snapshots}
\end{figure*}

\subsection{Initial conditions}
\emph{Stellar masses.} For the sake of simplicity, we model the stellar population as a two-component system consisting of low-mass stars of mass $m_\star = 0.2\,\Msol$, and of high-mass stars of mass $m_\star = 0.8\,\Msol$. Each sub-population contributes half of the total stellar mass $M_\star$. Consequently, in our models, the number of low-mass stars is four times higher than the number of high-mass stars. 
This choice of stellar masses is roughly guided by the \citet{Chabrier2003} present-day mass function, where half of the total stellar mass $M_\star$ is contributed by stars with masses below ${\sim}0.5\,\Msol$, with a median stellar mass of ${\sim}0.2\,\Msol$. The second half of the total stellar mass is contributed by stars with masses above ${\sim}0.5\,\Msol$, with a median stellar mass of ${\sim}0.8\,\Msol$. 

\emph{Stellar profiles.} We assume that, initially, the combined density profile of both low-mass and high-mass stars follows a spherical (3D) exponential profile, 
\begin{equation}
 \rho_\star(r) = \left[ M_\star / (8\pi r_\star^3) \right] ~ \exp(-r/r_\star)~, \label{Eq:StellarDensity}
\end{equation}
where $r_\star\approx\rh/2.67$ is the stellar scale radius. Initially, the stellar half-light radii $\rh$ coincide between the two sub-populations. We embed the stellar models deeply within the potential well of a smooth dark matter subhalo: $\rh / r_\mathrm{sub} = 1/500$, for a dark matter scale radius $r_\mathrm{sub}$ defined as follows. 

\emph{Dark matter profiles.} We model the (smooth) dark matter subhalo surrounding the stellar component and centered on it using a spherical \citet{hernquist1990} profile, with a total mass $M_\mathrm{sub}$ and a scale radius $r_\mathrm{sub}$,
\begin{equation}
 \rho_\mathrm{sub}(r) = \left[ M_\mathrm{sub} / (2\pi r_\mathrm{sub}^3) \right] ~  (r/ r_\mathrm{sub})^{-1}  (1 + r/r_\mathrm{sub})^{-3} ~. \label{Eq:Hernquist}
\end{equation}
This dark matter density profile is cuspy, i.e., $\diff \ln \rho_\mathrm{sub} / \diff \ln r \rightarrow -1$ for $r \rightarrow 0$.

\emph{$N$-body realizations.} We generate equilibrium $N$-body realizations of Eq.~\ref{Eq:StellarDensity} in the combined potential of the dark matter and stellar components using using the Eddington-inversion code \textsc{nbopy} \citep{EP20}, available online\footnote{\url{https://github.com/rerrani/nbopy}}. We assume that both stellar components have isotropic velocities in the initial conditions. The dark matter subhalo (Eq.~\ref{Eq:Hernquist}) is modeled as an analytical background potential. To reduce the impact of Poisson noise on our analysis, for each choice of dynamical-to-stellar mass ratio $\Upsilon_\mathrm{dyn}$, we generate $200$ realizations of our \texttt{UMa3/U1} model, $40$ for \texttt{Del1}, and $10$ for \texttt{Eri3}. 

\subsection{$N$-body code}
We compute the time evolution of our $N$-body models using \textsc{petar} \citep{Wang2020_PETAR}, a collisional $N$-body code, which in turn builds upon the slow-down arithmetic regularization package \textsc{sdar} \citep{Wang2020_SDAR} and the general-purpose library for particle simulations \textsc{fdps} \citep{Iwasawa2016, Namekata2018}. \textsc{petar} employs a fourth-order Hermite integrator to handle short-range forces, and a \citet{Barnes1986} tree for long-range ones. No force softening is used. The analytical Hernquist potential is included in the force calculations by making use of the code's \textsc{galpy} \citep{Bovy2015} interface. 

For the \texttt{UMa3/U1} models, we update the particle tree responsible for the long-range forces with a step size of $\Delta t_\mathrm{L} = 2^{-10} r_\mathrm{sub}^{3/2} (GM_\mathrm{sub})^{-1/2}$. With this choice, the period of a circular orbit at the initial half-light radius $\rh/r_\mathrm{sub}=1/500$ of a particle that is only subject to long-range forces (including the force due to the analytical dark matter potential) is resolved by ${\approx} 270$ steps for the models with $\Upsilon_\mathrm{dyn}=10$. To test for numerical convergence, we have decreased and increased this value by factors of 4, with no qualitative impact on our results. Following the recommendation of \citet[][see their Eq.~12 and 41]{Wang2020_PETAR}, we choose as reference radii for the separation of short- and long-range forces the values $r_\mathrm{in,ref} = \Delta t_\mathrm{L} \sigma_{1D}$ and $r_\mathrm{out,ref}=10\,r_\mathrm{in,ref}$, where by $\sigma_{1D}$ we denote the $N$-body system's velocity dispersion. Slow-down regularization through \textsc{sdar} is applied to particles below a radius $r_\mathrm{bin} = 0.8\,r_\mathrm{in,ref}$ (the default value in \textsc{petar}). To test for convergence, we have decreased this radius by a factor of 8, again with no qualitative impact on our results. 

For the \texttt{Del1} models, we use the same tree time step and reference radii as for \texttt{UMa3/U1}. For \texttt{Eri3}, we instead use a shorter tree time step of $\Delta t_\mathrm{L} = 2^{-12} r_\mathrm{sub}^{3/2} (GM_\mathrm{sub})^{-1/2}$, which results in better performance at that $N$-body particle number by reducing the number of particles within $r_\mathrm{in,ref}$.

\begin{figure*}
\centering
\includegraphics[width=6.6cm]{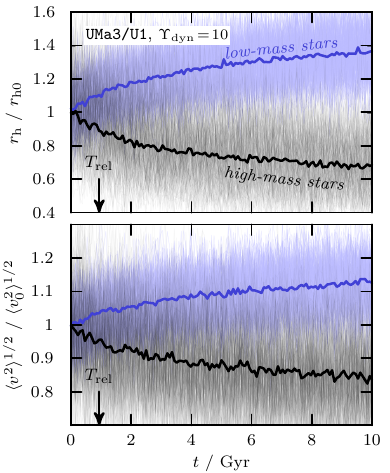}%
\includegraphics[width=5.7cm]{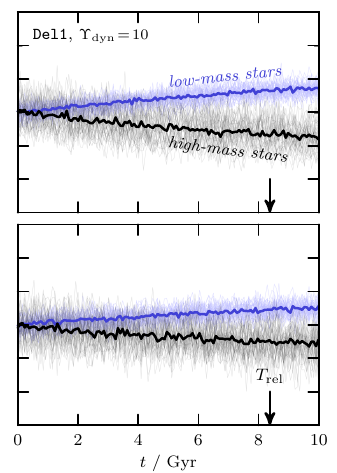}%
\includegraphics[width=5.7cm]{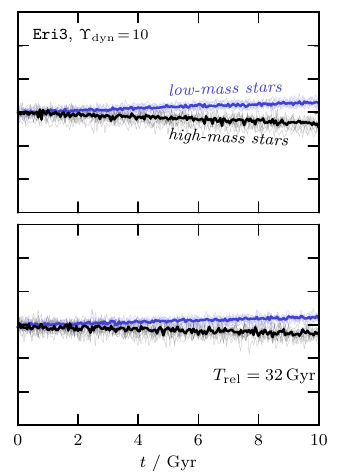}
\caption{As stellar mass segregation progresses, the half-light radius of the low-mass stars expands as the population heats up, whereas the population of high-mass stars contracts and cools down. The same systems are shown as in Fig.~\ref{fig:Snapshots}, with an initial dynamical-to-stellar mass ratio of $\Upsilon_\mathrm{dyn} = 10$. Thick blue (black) lines show the median time evolution of the low-mass (high-mass) stellar population computed from all $N$-body realizations (see text), whereas light blue (grey) lines show individual runs. Poisson noise widens the distribution of half-light radii and velocity dispersions, most notably for the case of \texttt{UMa3/U1} (left) with $N_\star=50$ stars.}
\label{fig:TimeEvolution}
\end{figure*}

\section{Simulation results}
\label{Sec:SimulationResults}
We can now turn our attention to the results of our $N$-body experiments. In Sec.~\ref{Sec:Static}, \ref{sec:Binaries} and \ref{Sec:DynamicalToStellarRatio}, we discuss simulations of \texttt{UMa3/U1}, \texttt{Del1} and \texttt{Eri3} where we embed the stellar component in a static dark matter halo. Then, in Sec.~\ref{Sec:Tides}, we model the effect of tides by allowing the underlying dark matter potential to evolve with time. 

\subsection{Simulations in a static dark matter halo}
\label{Sec:Static}
Fig.~\ref{fig:Snapshots} shows simulation snapshots of the \texttt{UMa3/U1} (top), \texttt{Del1} (center) and \texttt{Eri3} (bottom) models, evolved for $10\,\Gyr$ in a static dark matter halo. 
These models have an initial dynamical-to-stellar mass ratio of $\Upsilon_\mathrm{dyn}=10$ within the (initial) half-light radius. Individual low-mass (high-mass) stars are shown as light blue (grey) points, respectively. As time progresses, the population of low-mass stars expands, while the population of high-mass stars contracts: Even though the system is highly dark matter-dominated, the stellar population undergoes mass segregation\footnote{Mass segregation occurs as close encounters between stars provide a means for the exchange of energy. The system's tendency toward equipartition of energy results in low-mass stars preferentially moving to less-bound orbits (the low-mass population expands), whereas the high-mass stars move toward more tightly bound orbits (the high-mass population contracts), see, e.g., \citet[][chapter 1.3]{Spitzer1987book}.} driven by collisional relaxation. Blue and black circles in Fig.~\ref{fig:Snapshots} show the median half-light radii of the low-mass (high-mass) population, with the median computed from the sample of all $N$-body realizations. Mass segregation progresses fastest for \texttt{UMa3/U1} and slowest for \texttt{Eri3}, consistent with the relaxation times estimated in Fig.~\ref{fig:Trelax}.

The detailed time evolution of the models with $\Upsilon_\mathrm{dyn}=10$ is shown in Fig.~\ref{fig:TimeEvolution}. The top panels show the evolution of the (median) half-light radii of the populations of low-mass (blue) and high-mass (black) stars. For the case of \texttt{UMa3/U1}, after $10\,\Gyr$ of evolution, the half-light radius of the population of low-mass stars is twice as large as the half-light radius of high-mass stars. For \texttt{Del1}, the difference reduces to ${\sim}40$ per cent, and for \texttt{Eri3} to ${\sim}10$ per cent. The evolution of individual $N$-body realizations is shown in lighter shades in the background: For systems akin to \texttt{UMa3/U1}, Poisson noise driven by the low number of stars in the system substantially complicates a detection of mass segregation. The bottom panels of Fig.~\ref{fig:TimeEvolution} show the evolution of the (3D) stellar velocity dispersion $\langle v^2 \rangle = \sum v^2 / N_\star$. As the half-light radius of the population of low-mass stars expands, the velocity dispersion heats up. Vice versa, as the population of high-mass stars contracts, the population cools down. This behaviour is expected for gravitationally subdominant stellar components that are deeply embedded within a dark matter subhalo (see footnote~\ref{footnote:virial} for a calculation of the velocity dispersion predicted by the virial theorem, in good agreement with the simulation results, and \citealt{Penarrubia2025} for a detailed discussion of gravothermal expansion). Note that this behaviour differs from that of fully self-gravitating stellar systems devoid of dark matter, which typically heat up as they contract \citep[see, e.g.,][]{Lynden-BellWood1968, SpitzerThuan1972}.

\subsection{Dynamical formation and disruption of binaries}
\label{sec:Binaries}
As the population of high-mass stars contracts and cools, we note the dynamical formation and disruption of stellar binary systems in our simulations. We identify binaries as pairs of stars with Keplerian binding energy $E_\mathrm{bin} < 0$ and an orbital period $T_\mathrm{bin}$ that is shorter than the (circular) period $T_\mathrm{com}$ of the pair's center of mass within the dark matter subhalo. Expressed in terms of densities, this condition implies that the mean stellar density within the semimajor axis $a_\mathrm{bin}$ of a binary exceeds the mean density of the dark halo within a radius equal to that of the binary's center of mass , i.e $(m_i + m_j)/a_\mathrm{bin}^3 > M_\mathrm{sub}({<}r_\mathrm{com})/r_\mathrm{com}^3$, where by $m_i$ and $m_j$ we denote the masses of the two binary components.
We count the number of binaries at each snapshot in the simulation. For better statistics, we stack $2000$ realizations of the \texttt{UMa3/U1} model with an initial dynamical-to-stellar mass ratio of $\Upsilon_\mathrm{dyn} = 10$. Note that our initial conditions are created by drawing individual stars from the underlying distribution function. Any binaries present in the initial conditions just arise from this random sampling, resulting in a binary fraction that is substantially lower than current observational constraints\footnote{Spectroscopic studies suggest that the fraction of stellar binaries in dwarf galaxies is substantial: \citet{Spencer2018} infer a binary fraction of ${\sim}50$ per cent for Draco and ${\sim}80$ per cent in Ursa Minor.}. We adopt this modeling choice for the sake of simplicity.

\begin{figure}[tb]
\centering
\includegraphics[width=8.5cm]{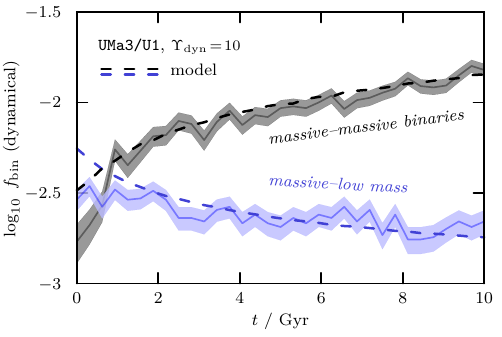}
\caption{As the population of high-mass stars contracts and cools, the number of dynamically formed binaries consisting of two massive stars increases. At the same time, as the population of low-mass stars expands and heats up, the number of dynamically formed binaries consisting of a massive and a low-mass star decreases. Shown is the fraction $f_\mathrm{bin}$ of high-mass stars that are in binary systems relative to the total number of high-mass stars. Simulation results are shown as shaded bands, computed from a sample of $2000$ $N$-body realizations of the \texttt{UMa3/U1} model. Dashed curves show the model predictions of Eq.~\ref{Eq:BinaryFraction}.}
\label{fig:Binaries}
\end{figure}

Fig.~\ref{fig:Binaries} shows the binary fraction $f_\mathrm{bin} \equiv N_{\mathrm{bin},i} / N_i$, defined here as the number $N_{\mathrm{bin},i}$ of high-mass stars that are in binary systems, normalized by the total number of high-mass stars $N_i$. A grey band shows the binary fraction considering only pairs of two high-mass stars, whereas the blue band corresponds to binaries that consist of a high-mass star and a low-mass star. As the population of high-mass stars contracts and cools due to mass segregation, the number of dynamically formed \emph{massive--massive} binaries increases. At the same time, as the population of low-mass stars expands and heats up, the number of \emph{massive--low\,mass} binaries drops. The stellar binaries in our simulations dynamically form and later dissolve again, consistent with the results of \citet{PetitHenon1986} who show that all gravitational captures (in three-body systems) are temporary configurations (see also \citealt{Penarrubia2023, Penarrubia2024}).

Some intuition in the processes driving the formation and disruption of dynamical binaries can be gained from the statistical theory of gravitational capture, developed to estimate the number of gravitationally trapped (massless) tracer particles around a point-mass perturber orbiting in a smooth dark matter halo \citep{Penarrubia2023}. In our collisional $N$-body simulations, all stars are massive particles, and complicated three- or many-body interaction between stars and the halo likely contribute to the formation and disruption processes. Nevertheless, as we will show in the following, the statistical theory of gravitational capture provides accurate estimates for the binary fractions found in our simulations. Building upon Eq.~4 in \citet[][]{Penarrubia2021}, we estimate the binary fraction through\footnote{We obtain our Eq.~\ref{Eq:BinaryFraction} from Eq.~4 of \citet[][]{Penarrubia2021} by approximating the mean number density of field stars trough $\bar n_j = 3 N_j / (8\pi r_{\mathrm{h}j}^3)$. We then substitute the mass of the perturber by the mass of the binary system, $m_i + m_j$. Finally, we compute the number of bound field stars within a radius $a_\mathrm{max} = r_{\mathrm{h}j} \left[  (m_i + m_j) /  M_\mathrm{sub}({<} r_{\mathrm{h}j})  \right]^{1/3} $.} %
\begin{equation}
 f_\mathrm{bin} \equiv \frac {N_{\mathrm{bin},i}}{N_i} \approx \frac{ 4 \sqrt{3}}{\sqrt{\pi}} \left[G\,(m_i+m_j)\, a_\mathrm{max}\right]^{3/2} \frac{N_j}{r_{{\mathrm{h}j}}^3 \langle v^2_j\rangle^{3/2} } 
 \label{Eq:BinaryFraction}
\end{equation}
where $N_i$ denotes the number and $m_i$ the mass of high-mass stars. Analogously, we denote by $N_j$, $m_j$, $r_{\mathrm{h}j}$ and $\langle v^2_j\rangle^{1/2}$ the number, mass, (3D) half-light radius and (3D) velocity dispersion of the field population. For \emph{massive--low\,mass} binaries, the field population is the population of low-mass stars. For \emph{massive--massive} binaries, the field population coincides with the population of massive stars, and we set $N_j = N_i -1$.
The radius $a_\mathrm{max}$ here is chosen to be the largest semimajor axis for which the binary identification criterion employed in the simulations holds, assuming a binary located at the half-light radius $r_{\mathrm{h}j}$.

From Eq.~\ref{Eq:BinaryFraction}, we see that the fraction of dynamically formed binaries scales with the (proxy) phase space density of the field population, $\propto N_j / (r_{\mathrm{h}j}^3 \langle v^2_j\rangle^{3/2})$. The latter increases for the case of \emph{massive--massive} binaries as the population of high-mass stars contracts and cools; hence, $f_\mathrm{bin}$ grows. Vice versa, the (proxy) phase space density of low-mass stars decreases as the population expands and heats up, and $f_\mathrm{bin}$ drops. Dashed curves in Fig.~\ref{fig:Binaries} show the evolution of $f_\mathrm{bin}$ predicted by Eq.~\ref{Eq:BinaryFraction}, which is in good agreement with the abundance of binaries identified in the $N$-body simulations.

In our simplified setup, most dynamically formed binaries are wide and thus contribute minimally to the systemic velocity dispersion. As a result, they may evade detection in spectroscopic surveys but can imprint a distinct signal in the stellar two-point correlation function \citep{LonghitanoBinggeli2010, Kervick2022, Safarzadeh2022}. Given that the statistics of wide binaries have been proposed as a sensitive probe of dark matter substructure \citep{PenarrubiaLudow2016, Cheyanne2025}, our simulation results, suggesting the dynamical formation and disruption of (wide) binaries in dark matter-dominated dwarf galaxies, emphasize the importance of detailed modeling of the dynamical processes that govern the abundance of wide binaries in these systems. 

\subsection{Sensitivity to the dynamical-to-stellar mass ratio }
\label{Sec:DynamicalToStellarRatio}
The models described in the previous sections assumed an initial dynamical-to-stellar mass ratio of $\Upsilon_\mathrm{dyn}=10$. However, Eq.~\ref{Eq:TrelaxUps} shows that the relaxation time, and hence the effects of collisional processes on the system, depend on the value of $\Upsilon_\mathrm{dyn}$. For $\Upsilon_\mathrm{dyn} \rightarrow \infty$, the system becomes collisionless, whereas for $\Upsilon_\mathrm{dyn} \rightarrow 1$, the dynamics are those of a classical star cluster. To study this dependence on $\Upsilon_\mathrm{dyn}$, we run a series of simulations, varying the initial dynamical-to-stellar mass ratio over the range $0.5 \leq \log_{10} \Upsilon_\mathrm{dyn} \leq 2.5$. Note that we only study models that are dark matter-dominated, as our $N$-body setup does not capture the dynamical effects of stars on the dark matter cusp, which would become more relevant as $\Upsilon_\mathrm{dyn}$ decreases (see, e.g., \citealt{ZhangAmaro2025}).

\begin{figure}[tb]
\centering
\includegraphics[width=8.5cm]{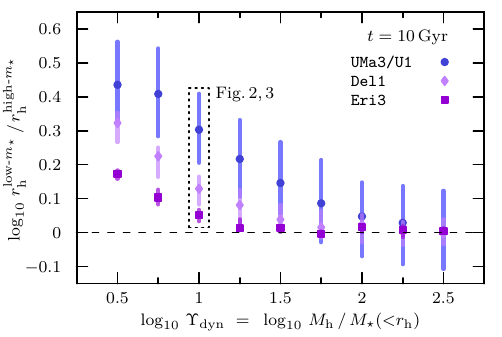}
\caption{Stellar mass segregation plays a substantial role in the dynamical evolution of systems akin to \texttt{UMa3/U1} if their dynamical-to-stellar mass ratios~$\Upsilon_\mathrm{dyn}$ are smaller than ${\sim}50$. Shown is the median ratio of half-light radii between the populations of low-mass and high-mass stars after $10\,\Gyr$ of evolution for different (initial) dynamical-to-stellar mass ratios $\Upsilon_\mathrm{dyn}$, computed over all $N$-body realizations (see text). Error bars span the $16^\mathrm{th}$ to $84^\mathrm{th}$ percentiles of the underlying distribution. }
\label{fig:MassToLight}
\end{figure}

In Fig.~\ref{fig:MassToLight}, we show the ratio between the half-light radii of the low-mass and high-mass stellar populations after $10\,\Gyr$ of evolution, for different values of the initial dynamical-to-stellar mass ratio $\Upsilon_\mathrm{dyn}$. As expected, for large values of $\Upsilon_\mathrm{dyn}$, there is no appreciable difference between the half-light radii of the two stellar populations: the system is collisionless. At the other extreme, for $\Upsilon_\mathrm{dyn} \sim 3$ and $10\,\Gyr$ of evolution, the stellar half-light radius of the low-mass population is three times, 2 times and 50 per cent larger than that of the population of low-mass stars for the cases of the \texttt{UMa3/U1}, \texttt{Del1} and \texttt{Eri3} models, respectively. Error bars span the $16^\mathrm{th}$ to $84^\mathrm{th}$ percentiles of the underlying distribution of $N$-body relizations. As before, Poisson noise renders any detection of mass segregation highly challenging for systems with a low number of member stars such as \texttt{UMa3/U1}.

\subsection{The effect of tides}
\label{Sec:Tides}
The simulation results discussed in Sec.~\ref{Sec:Static} assume a static dark matter potential surrounding the stellar populations. For the faint stellar systems \texttt{UMa3/U1}, \texttt{Del1} this assumption is unlikely to hold: located at galactocentric distances similar to that of the Sun, these systems will be subject to the tidal field of the Milky Way. 
For the case of \texttt{UMa3/U1} on an orbit with a pericentre of $r_\mathrm{peri}=13\,\kpc$, a dark matter halo hosting \texttt{UMa3/U1} could have been tidally stripped to $1/10^4$ of its original mass (see Fig.~7 of \citealt{ENSM24}).

\begin{figure}[tb]
 \centering
 \includegraphics[width=8.5cm]{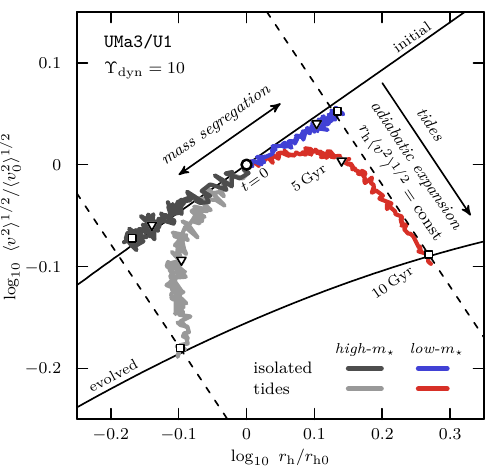}
 \caption{The effects of mass segregation on a stellar system are separable from tidal effects. The final ratio of half-light radii between the low-mass and high-mass populations is virtually unaffected by the adiabatic expansion of the stellar component in response to tides. Shown are the evolution of half-light radius and velocity dispersion of the \texttt{UMa3/U1} model, in isolation (blue and dark grey curves for the low- and high-mass populations, respectively), and for the case of a tidal field that slowly lowers the dark matter mass enclosed within the stellar component (red and light grey curves). Circles, triangles and squares mark simulation times of $0\,\Gyr$, $5\,\Gyr$ and $10\,\Gyr$. Black solid curves illustrate the potential landscape at the beginning of the simulation (top curve, ``initial'') and after $10\,\Gyr$ (bottom curve, ``evolved'') for the simulation with tides, see text for details. }
 \label{fig:Tides}
\end{figure}

To model the effect of tides, in the following, we will slowly decrease the mass of the surrounding dark matter halo and adjust its scale radius according to the empirical tidal evolutionary tracks for Hernquist models (see \citealt{EPW18} Table~A1 and Fig.~A1 for details). Specifically, we model the evolution of the subhalo total mass as $M_\mathrm{sub} / M_\mathrm{sub0} = \exp(-t/\tau) $ where $\tau$ is chosen so that over $10\,\Gyr$ of evolution, $M_\mathrm{sub} / M_\mathrm{sub0}$ decreases to $1/10^{4}$. The scale radius is adjusted through $r_\mathrm{sub}/r_\mathrm{sub0} \approx (M_\mathrm{sub} / M_\mathrm{sub0})^\beta$ with $\beta = 0.48$. 

Taking \texttt{UMa3/U1} as an example, Fig.~\ref{fig:Tides} shows the results of this experiment. A red curve shows the evolution of the (median) half-light radius $\rh$ and velocity dispersion $\langle v^2 \rangle^{1/2}$ of the population of low-mass stars, normalized to the respective initial values. A grey curve shows the equivalent evolution for the population of high-mass stars. As previously discussed, the stellar populations segregate in mass. Here, in addition, the lowering of the background dark matter potential drives an adiabatic expansion of the stellar components along curves of $\rh \langle v^2 \rangle^{1/2}=\text{const}$ (black dashed curves, see \citealt{EWRPN2025} for detailed discussion of this adiabatic expansion). Note that, in the size--velocity dispersion plane, the dynamical effects of tides and mass segregation are orthogonal: mass segregation results in an overall expansion and heating (contraction and cooling) of the population of low-mass (high-mass) stars, whereas tides drive an adiabatic expansion and cooling of both components. For the case of the population of high-mass stars, these two processes compete in driving the half-light radius of the stellar population. Crucially, after $10\,\Gyr$ of evolution, the ratio between the half-light radii of the low-mass and high-mass populations is virtually identical in the models that include adiabatic expansion through tides (red and grey lines), and in the models run in isolation (blue and dark grey lines). For the case of tidal fields that cause an adiabatic expansion of the stellar component within the power-law cusp of the underlying halo, the ratio of half-light radii shown in Fig.~\ref{fig:MassToLight} will hence hold independently of whether the system has experienced tides or not. This finding is consistent with the analytical estimate of the relaxation time in Eq.~\ref{Eq:TrelaxMhrh}: for constant $N_\star$ and $m_\star$, the relaxation timescales as $T_\mathrm{rel} \propto (\Mh \rh)^{3/2}$. The product of enclosed mass and half-light radius is approximately conserved during adiabatic expansion, $\Mh \rh \propto \langle v^2 \rangle^{1/2} \rh \approx\text{const}$ (see, e.g., \citealt{EWRPN2025} for details). Hence, the relaxation time (Eq.~\ref{Eq:TrelaxMhrh}) and the binary fraction (Eq.~\ref{Eq:BinaryFraction}) remain virtually unaffected by the adiabatic expansion of the stellar components.

To illustrate the potential landscape and to provide further intuition for the expected evolution in the size--velocity dispersion plane, black curves show the velocity dispersion of a tracer population subject to the combined potential of dark matter and stars, as computed%
\footnote{Velocity dispersions are additive, and we can compute separately the dispersions expected for a stellar component due to (a) the potential of another stellar component, (b) its self gravity, and (c) the Hernquist dark matter halo. In each case, we will use Eq.~3 of \citet{EPW18} for the calculation of the velocity dispersion. The velocity dispersion of a (massless) exponential stellar tracer (Eq.~\ref{Eq:StellarDensity}) with scale radius $r_i$ in the potential of another stellar component of mass $M_j$ and scale radius $r_j$ reads
\begin{equation}
 \langle v^2_\text{a} \rangle = \frac{GM_j}{2 r_i} \left( 1 + 4 r_j / r_i \right) \left( 1 + r_j / r_i \right)^{-4}
\end{equation}
which for the self-gravitating case, $r_i = r_j$ and $M_i = M_j$ reduces to $\langle v^2_\text{b} \rangle = (5/96)~ GM_i/r_i$.
For an exponential stellar tracer with scale radius $r_i$ embedded in a Hernquist potential (Eq.~\ref{Eq:Hernquist}), writing for short $x=r_\mathrm{sub}/r_i$, we find
\begin{equation}
 \langle v^2_\text{c} \rangle = \frac{GM_\mathrm{sub}}{2 r_i} \left[  2 \!-\! (1 \!+\! x)^2 \!-\!  x^2  (x \!+\! 3) \exp(x)  \mathrm{Ei}(-x) \right] 
 \label{Eq:SigmaExpInHernquist}
\end{equation}
which for deeply embedded systems, $x \gg 1$, approaches the power law $\langle v^2_\mathrm{c} \rangle \approx 3 GM_\mathrm{sub} r_i/ r_\mathrm{sub}^2$. Note that for large values of $x$, numerical evaluations of Eq.~\ref{Eq:SigmaExpInHernquist} may be facilitated by expressing the product $ \exp(x) \mathrm{Ei}(-x) $ by its approximation as a Laurent series: $ - x^{-1} + x^{-2} - 2x^{-3} + 6x^{-4}-24x^{-5}+ \mathcal{O}(x^{-6}) $.\label{footnote:virial}}
from the virial theorem (see, e.g., \citealt{Amorisco2012}; \citealt{EPW18}). This simple calculation accurately predicts the velocity dispersion of the mass-segregated populations.

The model for tides employed here does not account for any stellar mass loss, but merely models the response of the stars to the evolving background potential. This is a modeling choice motivated by the fact that the asymptotic remnant state \citep{EN21} of a cold dark matter halo on the orbit of \texttt{UMa3/U1} has a tidal radius that is substantially larger than the half-light radius of \texttt{UMa3/U1}. For stronger tidal fields that result in the tidal stripping of stars, this assumption will not hold. As mass segregation drives low-mass stars to less bound orbits, and high-mass stars to more bound ones, in a mass-segregated system, tides would first strip the population of low-mass stars. This could result in the existence of a population of dark matter subhalos that hosts high-mass stars or their remnants at their centers: black holes surrounded by dark matter subhalos. Their mergers would in turn facilitate the formation of massive black holes in the centres of dark matter-dominated systems. This will be explored in future contributions.

\section{Conclusions}
\label{sec:conclusions}
\emph{Summary.} In the present work, we show that effects of collisional relaxation may play a substantial role in the dynamical evolution of a stellar component even in a dark matter-dominated system. This is of particular relevance for old stellar systems with short crossing times, where small collisional perturbations can accumulate over the course of several gigayears. We show that for such systems, collisional relaxation drives stellar mass segregation and the dynamical formation of binaries even in the presence of a gravitationally dominant smooth dark matter component. Our results hence call for caution when using stellar mass segregation as a litmus test for the absence of dark matter in ambiguous stellar clusters \citep{Kim2015, Baumgardt2022, Simon2024, Zaremba2025} and tidal streams \citep{ENIM2022}. Detailed modeling of the relaxation timescale and the Poisson noise floor is required to put constraints on the dark matter content of a stellar system through the observable signatures of mass segregation.

\emph{Caveats.} Our models make various simplifying assumptions. Most crucially, the stellar populations are modeled as a two-component system of high- and low-mass stars, the two initially sharing the same half-light radius. The relaxation time of a stellar system depends sensitively on its stellar mass function and the abundance of high-mass stars. Our models do not include high-mass stellar remnants, which may constitute a source of additional collisional perturbations that could amplify and speed up mass segregation. In that regard we believe our modeling choices to be conservative, putting a lower bound on the amount of mass segregation that is to be expected in the presence of a smooth dark matter subhalo. 

Furthermore, our models are tailored to describe systems that remain dark matter-dominated throughout their evolution, and neglect the dynamical effects of the stars on the dark matter distribution. The same fluctuating tidal field that drives stellar mass segregation is likely to affect the dark matter as well, particularly in systems that are initially or become baryon-dominated. A detailed study of this effect is computationally expensive and beyond the scope of the current paper. 

\emph{Outlook.} The models developed in this work are motivated by the recent discovery of ambiguous stellar systems at the interface of the dwarf galaxy and globular cluster regimes; nevertheless they can also find application in understanding the dynamical processes in the central regions of dwarf spheroidal and ultrafaint galaxies. While their half-mass relaxation times may exceed the age of the Universe, mass segregation plausibly still plays a role in their centres, where dynamical timescales and stellar densities are similar to those of the systems studied here. 
In addition to the effects of dynamical friction by the dark matter component, stellar mass segregation may further enhance the clustering of massive stars and their remnants in the centres of dwarf galaxies, plausibly playing a role in setting their merger rates and a potential accompanying gravitational wave signal. This in turn would facilitate the formation of massive black holes in the centres of dark matter-dominated dwarf galaxies (see, e.g., \citealt{Bustamante-Rosell2021} for the case of Leo~I). A quantitative analysis of this effect requires a detailed modeling of the stellar mass function beyond the two-component setup used in the present work, as well as a detailed modeling of the dynamical response of the dark matter cusp, which we defer to subsequent study.

\section*{Acknowledgements}
The authors would like to thank Anna Lisa Varri, Rodrigo Ibata, Giacomo Monari, and Josh Simon for stimulating discussions. RE and MW acknowledge support from the National Science Foundation (NSF) grant AST-2206046. Support for program JWST-AR-02352.001-A was provided by NASA through a grant from the Space Telescope Science Institute, which is operated by the Association of Universities for Research in Astronomy, Inc., under NASA contract NAS 5-03127. This material is based upon work supported by the National Aeronautics and Space Administration under Grant/Agreement No. 80NSSC24K0084 as part of the Roman Large Wide Field Science program funded through ROSES call NNH22ZDA001N-ROMAN.

\appendix

\section{Supplementary animated figures}
\label{Appendix:Animations}

\begin{figure}[b]
 \centering
 \fbox{\includegraphics[width=8.5cm]{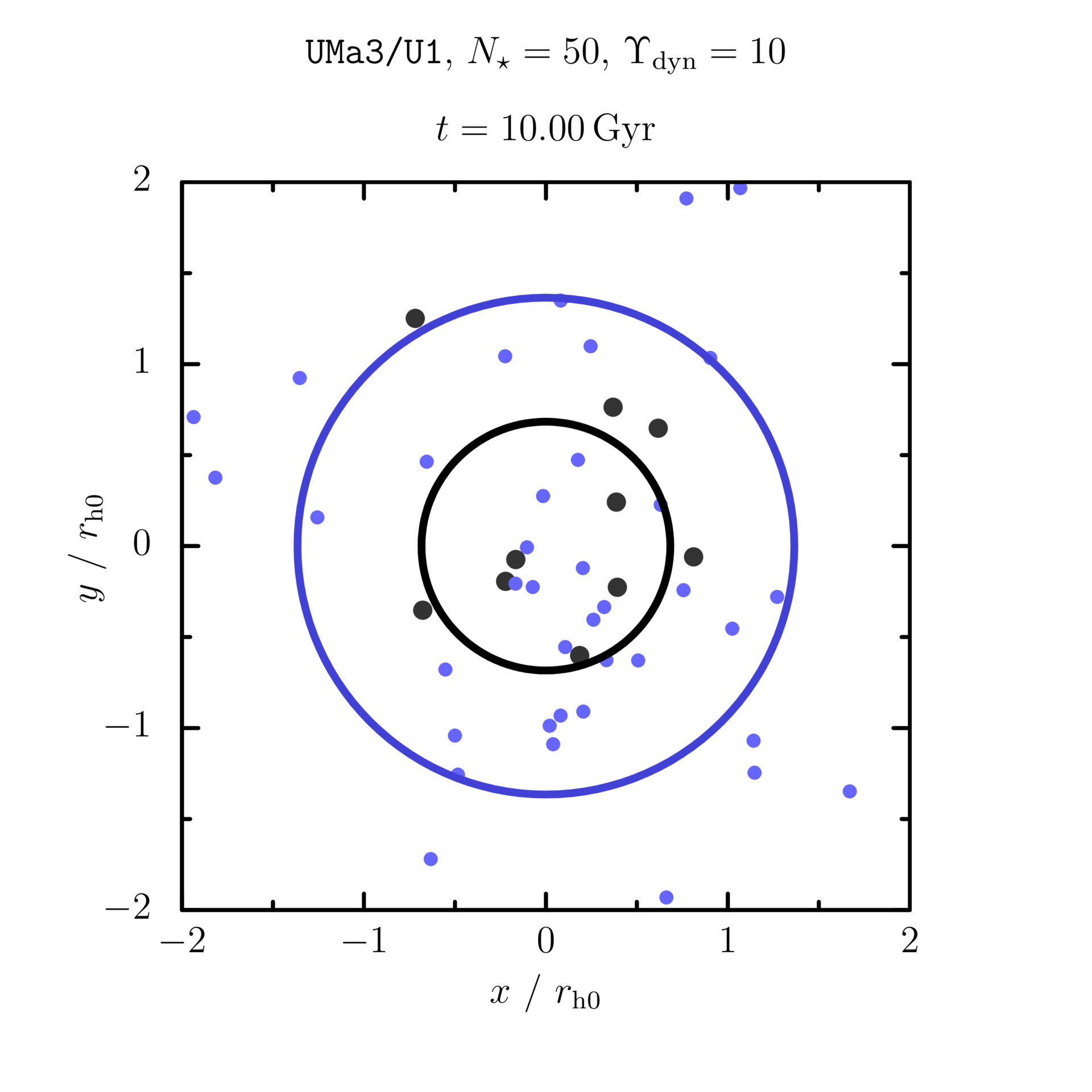}}
 \caption{Animated figure supplementing the top panel of Fig.~\ref{fig:Snapshots}. The video shows the time evolution of a population of high- and low-mass stars (black and blue points, respectively) in the presence of a smooth dark matter potential. Circles indicate the median half-light radii measured from a sample of $N$-body realizations (see Sec.~\ref{Sec:Static}). File size $4.9\,\mathrm{MB}$, dimensions $1920\times1920\,\mathrm{px}^2$, duration $40\,\mathrm{s}$. The animation shows the evolution from $0$ to $10\,\Gyr$. The playback speed is increased between seconds $14$ ($0.5\,\Gyr$) and $26$ ($9.5\,\Gyr$) for the sake of brevity. The animated figure is available in the \href{https://arxiv.org/src/2505.22717v2/anc}{arXiv ancillary files}.}
 \label{fig:ani_snapshots}
\end{figure}

\begin{figure}[t]
 \centering
 \fbox{\includegraphics[width=8.5cm]{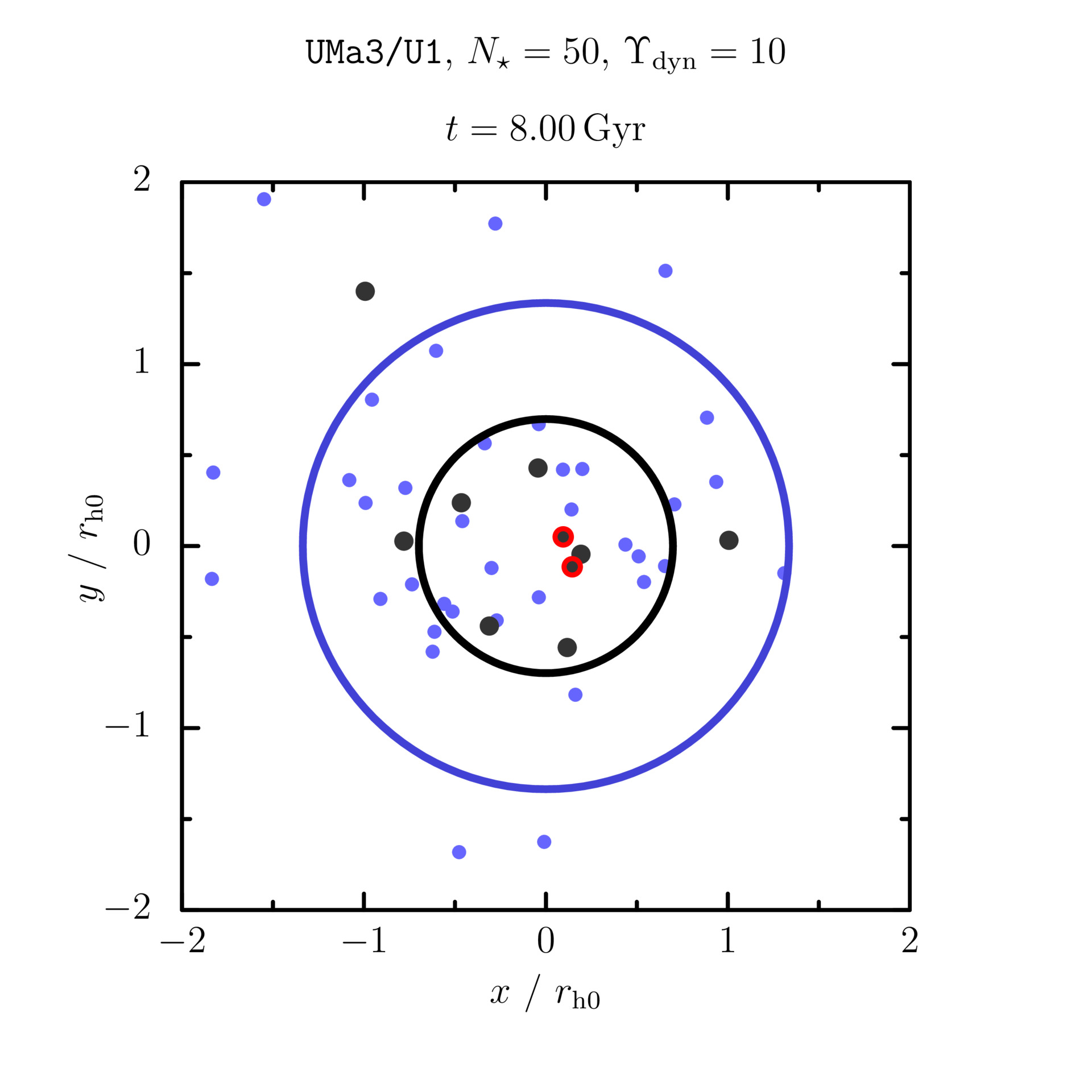}}
 \caption{Like Fig.~\ref{fig:ani_snapshots}, but highlighting the dynamical formation and subsequent disruption of a stellar binary (red circles). File size $1.8\,\mathrm{MB}$, dimensions $1920\times1920\,\mathrm{px}^2$, duration $28\,\mathrm{s}$. The animation shows the evolution from $7.75$ to $8.25\,\Gyr$. The animated figure is available in the \href{https://arxiv.org/src/2505.22717v2/anc}{arXiv ancillary files}.}
 \label{fig:ani_binary}
\end{figure}

This appendix contains brief descriptions of the animated figures \ref{fig:ani_snapshots} and \ref{fig:ani_binary}, available in the \href{https://arxiv.org/src/2505.22717v2/anc}{arXiv ancillary files}.

\newpage
\section*{References}
\bibliographystyle{aasjournal-hyperref}
\bibliography{relax}

\begin{thebibliography}{}
\expandafter\ifx\csname natexlab\endcsname\relax\def\natexlab#1{#1}\fi
\providecommand{\url}[1]{\href{#1}{#1}}
\providecommand{\dodoi}[1]{doi:~\href{http://doi.org/#1}{\nolinkurl{#1}}}
\providecommand{\doeprint}[1]{\href{http://ascl.net/#1}{\nolinkurl{http://ascl.net/#1}}}
\providecommand{\doarXiv}[1]{\href{https://arxiv.org/abs/#1}{\nolinkurl{https://arxiv.org/abs/#1}}}

\bibitem[{{Aarseth}(1999)}]{Aarseth1999}
{Aarseth}, S.~J. 1999,
  \href{http://dx.doi.org/10.1086/316455}{\color{magenta}\pasp},
  \href{https://ui.adsabs.harvard.edu/abs/1999PASP..111.1333A}{\color{blue}111},
  1333

\bibitem[{{Amorisco} \& {Evans}(2012)}]{Amorisco2012}
{Amorisco}, N.~C., \& {Evans}, N.~W. 2012,
  \href{http://dx.doi.org/10.1111/j.1365-2966.2011.19684.x}{\color{magenta}\mnras},
  \href{http://adsabs.harvard.edu/abs/2012MNRAS.419..184A}{\color{blue}419},
  184

\bibitem[{{Barnes} \& {Hut}(1986)}]{Barnes1986}
{Barnes}, J., \& {Hut}, P. 1986,
  \href{http://dx.doi.org/10.1038/324446a0}{\color{magenta}\nat},
  \href{https://ui.adsabs.harvard.edu/abs/1986Natur.324..446B}{\color{blue}324},
  446

\bibitem[{{Bastian} \& {Lardo}(2018)}]{BastianLardo2018}
{Bastian}, N., \& {Lardo}, C. 2018,
  \href{http://dx.doi.org/10.1146/annurev-astro-081817-051839}{\color{magenta}\araa},
  \href{https://ui.adsabs.harvard.edu/abs/2018ARA&A..56...83B}{\color{blue}56},
  83

\bibitem[{{Baumgardt} {et~al.}(2022){Baumgardt}, {Faller}, {Meinhold},
  {McGovern-Greco}, \& {Hilker}}]{Baumgardt2022}
{Baumgardt}, H., {Faller}, J., {Meinhold}, N., {McGovern-Greco}, C., \&
  {Hilker}, M. 2022,
  \href{http://dx.doi.org/10.1093/mnras/stab3629}{\color{magenta}\mnras},
  \href{https://ui.adsabs.harvard.edu/abs/2022MNRAS.510.3531B}{\color{blue}510},
  3531

\bibitem[{{Bovy}(2015)}]{Bovy2015}
{Bovy}, J. 2015,
  \href{http://dx.doi.org/10.1088/0067-0049/216/2/29}{\color{magenta}\apjs},
  \href{https://ui.adsabs.harvard.edu/abs/2015ApJS..216...29B}{\color{blue}216},
  29

\bibitem[{{Bruce} {et~al.}(2023){Bruce}, {Li}, {Pace}, {Heiger}, {Song}, \&
  {Simon}}]{Bruce2023}
{Bruce}, J., {Li}, T.~S., {Pace}, A.~B., {et~al.} 2023,
  \href{http://dx.doi.org/10.3847/1538-4357/acc943}{\color{magenta}\apj},
  \href{https://ui.adsabs.harvard.edu/abs/2023ApJ...950..167B}{\color{blue}950},
  167

\bibitem[{{Bustamante-Rosell} {et~al.}(2021){Bustamante-Rosell}, {Noyola},
  {Gebhardt}, {Fabricius}, {Mazzalay}, {Thomas}, \&
  {Zeimann}}]{Bustamante-Rosell2021}
{Bustamante-Rosell}, M.~J., {Noyola}, E., {Gebhardt}, K., {et~al.} 2021,
  \href{http://dx.doi.org/10.3847/1538-4357/ac0c79}{\color{magenta}\apj},
  \href{https://ui.adsabs.harvard.edu/abs/2021ApJ...921..107B}{\color{blue}921},
  107

\bibitem[{{Buttry} {et~al.}(2022){Buttry}, {Pace}, {Koposov}, {Walker},
  {Caldwell}, {Kirby}, {Martin}, {Mateo}, {Olszewski}, {Starkenburg},
  {Badenes}, \& {Daher}}]{Buttry2022}
{Buttry}, R., {Pace}, A.~B., {Koposov}, S.~E., {et~al.} 2022,
  \href{http://dx.doi.org/10.1093/mnras/stac1441}{\color{magenta}\mnras},
  \href{https://ui.adsabs.harvard.edu/abs/2022MNRAS.514.1706B}{\color{blue}514},
  1706

\bibitem[{{Cerny} {et~al.}(2023{\natexlab{a}}){Cerny},
  {Mart{\'\i}nez-V{\'a}zquez}, {Drlica-Wagner}, {Pace}, {Mutlu-Pakdil}, {Li},
  {Riley}, {Crnojevi{\'c}}, {Bom}, {Carballo-Bello}, {Carlin}, {Chiti}, {Choi},
  {Collins}, {Darragh-Ford}, {Ferguson}, {Geha}, {Mart{\'\i}nez-Delgado},
  {Massana}, {Mau}, {Medina}, {Mu{\~n}oz}, {Nadler}, {No{\"e}l}, {Olsen},
  {Pieres}, {Sakowska}, {Simon}, {Stringfellow}, {Tollerud}, {Vivas}, {Walker},
  {Wechsler}, \& {Delve Collaboration}}]{Cerny22}
{Cerny}, W., {Mart{\'\i}nez-V{\'a}zquez}, C.~E., {Drlica-Wagner}, A., {et~al.}
  2023{\natexlab{a}},
  \href{http://dx.doi.org/10.3847/1538-4357/acdd78}{\color{magenta}\apj},
  \href{https://ui.adsabs.harvard.edu/abs/2023ApJ...953....1C}{\color{blue}953},
  1

\bibitem[{{Cerny} {et~al.}(2023{\natexlab{b}}){Cerny}, {Simon}, {Li},
  {Drlica-Wagner}, {Pace}, {Mart{\'\i}nez-V{\'a}zquez}, {Riley},
  {Mutlu-Pakdil}, {Mau}, {Ferguson}, {Erkal}, {Munoz}, {Bom}, {Carlin},
  {Carollo}, {Choi}, {Ji}, {Manwadkar}, {Mart{\'\i}nez-Delgado}, {Miller},
  {No{\"e}l}, {Sakowska}, {Sand}, {Stringfellow}, {Tollerud}, {Vivas},
  {Carballo-Bello}, {Hernandez-Lang}, {James}, {Nidever}, {Castellon}, {Olsen},
  {Zenteno}, \& {Delve Collaboration}}]{Cerny23}
{Cerny}, W., {Simon}, J.~D., {Li}, T.~S., {et~al.} 2023{\natexlab{b}},
  \href{http://dx.doi.org/10.3847/1538-4357/aca1c3}{\color{magenta}\apj},
  \href{https://ui.adsabs.harvard.edu/abs/2023ApJ...942..111C}{\color{blue}942},
  111

\bibitem[{{Chabrier}(2003)}]{Chabrier2003}
{Chabrier}, G. 2003,
  \href{http://dx.doi.org/10.1086/376392}{\color{magenta}\pasp},
  \href{https://ui.adsabs.harvard.edu/abs/2003PASP..115..763C}{\color{blue}115},
  763

\bibitem[{{Conn} {et~al.}(2018){Conn}, {Jerjen}, {Kim}, \&
  {Schirmer}}]{Conn2018}
{Conn}, B.~C., {Jerjen}, H., {Kim}, D., \& {Schirmer}, M. 2018,
  \href{http://dx.doi.org/10.3847/1538-4357/aa9eda}{\color{magenta}\apj},
  \href{https://ui.adsabs.harvard.edu/abs/2018ApJ...852...68C}{\color{blue}852},
  68

\bibitem[{{Crnogor{\v{c}}evi{\'c}} \& {Linden}(2024)}]{Crnogorcevic2024}
{Crnogor{\v{c}}evi{\'c}}, M., \& {Linden}, T. 2024,
  \href{http://dx.doi.org/10.1103/PhysRevD.109.083018}{\color{magenta}\prd},
  \href{https://ui.adsabs.harvard.edu/abs/2024PhRvD.109h3018C}{\color{blue}109},
  083018

\bibitem[{{Devlin} {et~al.}(2025){Devlin}, {Baumgardt}, \&
  {Sweet}}]{Devlin2025}
{Devlin}, S., {Baumgardt}, H., \& {Sweet}, S.~M. 2025,
  \href{http://dx.doi.org/10.1093/mnras/staf572}{\color{magenta}\mnras},
  \href{https://ui.adsabs.harvard.edu/abs/2025MNRAS.539.2485D}{\color{blue}539},
  2485

\bibitem[{{Errani} {et~al.}(2024{\natexlab{a}}){Errani}, {Ibata}, {Navarro},
  {Pe{\~n}arrubia}, \& {Walker}}]{EINPW2024}
{Errani}, R., {Ibata}, R., {Navarro}, J.~F., {Pe{\~n}arrubia}, J., \& {Walker},
  M.~G. 2024{\natexlab{a}},
  \href{http://dx.doi.org/10.3847/1538-4357/ad402d}{\color{magenta}\apj},
  \href{https://ui.adsabs.harvard.edu/abs/2024ApJ...968...89E}{\color{blue}968},
  89

\bibitem[{{Errani} \& {Navarro}(2021)}]{EN21}
{Errani}, R., \& {Navarro}, J.~F. 2021,
  \href{http://dx.doi.org/10.1093/mnras/stab1215}{\color{magenta}\mnras},
  \href{https://ui.adsabs.harvard.edu/abs/2021MNRAS.505...18E}{\color{blue}505},
  18

\bibitem[{{Errani} {et~al.}(2024{\natexlab{b}}){Errani}, {Navarro}, {Smith}, \&
  {McConnachie}}]{ENSM24}
{Errani}, R., {Navarro}, J.~F., {Smith}, S. E.~T., \& {McConnachie}, A.~W.
  2024{\natexlab{b}},
  \href{http://dx.doi.org/10.3847/1538-4357/ad2267}{\color{magenta}\apj},
  \href{https://ui.adsabs.harvard.edu/abs/2024ApJ...965...20E}{\color{blue}965},
  20

\bibitem[{{Errani} \& {Pe{\~n}arrubia}(2020)}]{EP20}
{Errani}, R., \& {Pe{\~n}arrubia}, J. 2020,
  \href{http://dx.doi.org/10.1093/mnras/stz3349}{\color{magenta}\mnras},
  \href{https://ui.adsabs.harvard.edu/abs/2020MNRAS.491.4591E}{\color{blue}491},
  4591

\bibitem[{{Errani} {et~al.}(2018){Errani}, {Pe{\~n}arrubia}, \&
  {Walker}}]{EPW18}
{Errani}, R., {Pe{\~n}arrubia}, J., \& {Walker}, M.~G. 2018,
  \href{http://dx.doi.org/10.1093/mnras/sty2505}{\color{magenta}\mnras},
  \href{https://ui.adsabs.harvard.edu/abs/2018MNRAS.481.5073E}{\color{blue}481},
  5073

\bibitem[{{Errani} {et~al.}(2025){Errani}, {Walker}, {Rozier},
  {Pe{\~n}arrubia}, \& {Navarro}}]{EWRPN2025}
{Errani}, R., {Walker}, M.~G., {Rozier}, S., {Pe{\~n}arrubia}, J., \&
  {Navarro}, J.~F. 2025,
  \href{http://dx.doi.org/10.3847/1538-4357/adfa27}{\color{magenta}\apj},
  \href{https://ui.adsabs.harvard.edu/abs/2025ApJ...992..162E}{\color{blue}992},
  162

\bibitem[{{Errani} {et~al.}(2022){Errani}, {Navarro}, {Ibata}, {Martin},
  {Yuan}, {Aguado}, {Bonifacio}, {Caffau}, {Gonz{\'a}lez Hern{\'a}ndez},
  {Malhan}, {S{\'a}nchez-Janssen}, {Sestito}, {Starkenburg}, {Thomas}, \&
  {Venn}}]{ENIM2022}
{Errani}, R., {Navarro}, J.~F., {Ibata}, R., {et~al.} 2022,
  \href{http://dx.doi.org/10.1093/mnras/stac1516}{\color{magenta}\mnras},
  \href{https://ui.adsabs.harvard.edu/abs/2022MNRAS.514.3532E}{\color{blue}514},
  3532

\bibitem[{{Fellhauer} {et~al.}(2000){Fellhauer}, {Kroupa}, {Baumgardt}, {Bien},
  {Boily}, {Spurzem}, \& {Wassmer}}]{Fellhauer2000}
{Fellhauer}, M., {Kroupa}, P., {Baumgardt}, H., {et~al.} 2000,
  \href{http://dx.doi.org/10.1016/S1384-1076(00)00032-4}{\color{magenta}NA},
  \href{http://adsabs.harvard.edu/abs/2000NewA....5..305F}{\color{blue}5}, 305

\bibitem[{{Fu} {et~al.}(2023){Fu}, {Weisz}, {Starkenburg}, {Martin}, {Savino},
  {Boylan-Kolchin}, {C{\^o}t{\'e}}, {Dolphin}, {Ji}, {Longeard}, {Mateo},
  {Patel}, \& {Sandford}}]{Fu2023}
{Fu}, S.~W., {Weisz}, D.~R., {Starkenburg}, E., {et~al.} 2023,
  \href{http://dx.doi.org/10.3847/1538-4357/ad0030}{\color{magenta}\apj},
  \href{https://ui.adsabs.harvard.edu/abs/2023ApJ...958..167F}{\color{blue}958},
  167

\bibitem[{{Graham} \& {Ramani}(2024)}]{Graham2024}
{Graham}, P.~W., \& {Ramani}, H. 2024,
  \href{http://dx.doi.org/10.1103/PhysRevD.110.075011}{\color{magenta}\prd},
  \href{https://ui.adsabs.harvard.edu/abs/2024PhRvD.110g5011G}{\color{blue}110},
  075011

\bibitem[{{Gratton} {et~al.}(2012){Gratton}, {Carretta}, \&
  {Bragaglia}}]{Gratton2012}
{Gratton}, R.~G., {Carretta}, E., \& {Bragaglia}, A. 2012,
  \href{http://dx.doi.org/10.1007/s00159-012-0050-3}{\color{magenta}\aapr},
  \href{https://ui.adsabs.harvard.edu/abs/2012A&ARv..20...50G}{\color{blue}20},
  50

\bibitem[{{Hernquist}(1990)}]{hernquist1990}
{Hernquist}, L. 1990,
  \href{http://dx.doi.org/10.1086/168845}{\color{magenta}ApJ},
  \href{http://adsabs.harvard.edu/abs/1990ApJ...356..359H}{\color{blue}356},
  359

\bibitem[{{Iwasawa} {et~al.}(2016){Iwasawa}, {Tanikawa}, {Hosono}, {Nitadori},
  {Muranushi}, \& {Makino}}]{Iwasawa2016}
{Iwasawa}, M., {Tanikawa}, A., {Hosono}, N., {et~al.} 2016,
  \href{http://dx.doi.org/10.1093/pasj/psw053}{\color{magenta}\pasj},
  \href{https://ui.adsabs.harvard.edu/abs/2016PASJ...68...54I}{\color{blue}68},
  54

\bibitem[{{Ji} {et~al.}(2019){Ji}, {Simon}, {Frebel}, {Venn}, \&
  {Hansen}}]{Ji2019_Gru1}
{Ji}, A.~P., {Simon}, J.~D., {Frebel}, A., {Venn}, K.~A., \& {Hansen}, T.~T.
  2019, \href{http://dx.doi.org/10.3847/1538-4357/aaf3bb}{\color{magenta}\apj},
  \href{https://ui.adsabs.harvard.edu/abs/2019ApJ...870...83J}{\color{blue}870},
  83

\bibitem[{{Kervick} {et~al.}(2022){Kervick}, {Walker}, {Pe{\~n}arrubia}, \&
  {Koposov}}]{Kervick2022}
{Kervick}, C., {Walker}, M.~G., {Pe{\~n}arrubia}, J., \& {Koposov}, S.~E. 2022,
  \href{http://dx.doi.org/10.3847/1538-4357/ac5b5f}{\color{magenta}\apj},
  \href{https://ui.adsabs.harvard.edu/abs/2022ApJ...929...77K}{\color{blue}929},
  77

\bibitem[{{Kim} {et~al.}(2015){Kim}, {Jerjen}, {Milone}, {Mackey}, \& {Da
  Costa}}]{Kim2015}
{Kim}, D., {Jerjen}, H., {Milone}, A.~P., {Mackey}, D., \& {Da Costa}, G.~S.
  2015,
  \href{http://dx.doi.org/10.1088/0004-637X/803/2/63}{\color{magenta}\apj},
  \href{https://ui.adsabs.harvard.edu/abs/2015ApJ...803...63K}{\color{blue}803},
  63

\bibitem[{{Kirby} {et~al.}(2017){Kirby}, {Cohen}, {Simon}, {Guhathakurta},
  {Thygesen}, \& {Duggan}}]{Kirby2017}
{Kirby}, E.~N., {Cohen}, J.~G., {Simon}, J.~D., {et~al.} 2017,
  \href{http://dx.doi.org/10.3847/1538-4357/aa6570}{\color{magenta}\apj},
  \href{https://ui.adsabs.harvard.edu/abs/2017ApJ...838...83K}{\color{blue}838},
  83

\bibitem[{{Longeard} {et~al.}(2018){Longeard}, {Martin}, {Starkenburg},
  {Ibata}, {Collins}, {Geha}, {Laevens}, {Rich}, {Aguado}, {Arentsen},
  {Carlberg}, {C{\^o}t{\'e}}, {Hill}, {Jablonka}, {Gonz{\'a}lez Hern{\'a}ndez},
  {Navarro}, {S{\'a}nchez-Janssen}, {Tolstoy}, {Venn}, \&
  {Youakim}}]{Longeard2018}
{Longeard}, N., {Martin}, N., {Starkenburg}, E., {et~al.} 2018,
  \href{http://dx.doi.org/10.1093/mnras/sty1986}{\color{magenta}\mnras},
  \href{https://ui.adsabs.harvard.edu/abs/2018MNRAS.480.2609L}{\color{blue}480},
  2609

\bibitem[{{Longhitano} \& {Binggeli}(2010)}]{LonghitanoBinggeli2010}
{Longhitano}, M., \& {Binggeli}, B. 2010,
  \href{http://dx.doi.org/10.1051/0004-6361/200913109}{\color{magenta}\aap},
  \href{https://ui.adsabs.harvard.edu/abs/2010A&A...509A..46L}{\color{blue}509},
  A46

\bibitem[{{Lynden-Bell} \& {Wood}(1968)}]{Lynden-BellWood1968}
{Lynden-Bell}, D., \& {Wood}, R. 1968,
  \href{http://dx.doi.org/10.1093/mnras/138.4.495}{\color{magenta}\mnras},
  \href{https://ui.adsabs.harvard.edu/abs/1968MNRAS.138..495L}{\color{blue}138},
  495

\bibitem[{{Martin} {et~al.}(2016{\natexlab{a}}){Martin}, {Geha}, {Ibata},
  {Collins}, {Laevens}, {Bell}, {Rix}, {Ferguson}, {Chambers}, {Wainscoat}, \&
  {Waters}}]{Martin2016_Dra2}
{Martin}, N.~F., {Geha}, M., {Ibata}, R.~A., {et~al.} 2016{\natexlab{a}},
  \href{http://dx.doi.org/10.1093/mnrasl/slw013}{\color{magenta}\mnras},
  \href{https://ui.adsabs.harvard.edu/abs/2016MNRAS.458L..59M}{\color{blue}458},
  L59

\bibitem[{{Martin} {et~al.}(2016{\natexlab{b}}){Martin}, {Ibata}, {Collins},
  {Rich}, {Bell}, {Ferguson}, {Laevens}, {Rix}, {Chapman}, \&
  {Koch}}]{Martin2016_Tri2}
{Martin}, N.~F., {Ibata}, R.~A., {Collins}, M. L.~M., {et~al.}
  2016{\natexlab{b}},
  \href{http://dx.doi.org/10.3847/0004-637X/818/1/40}{\color{magenta}\apj},
  \href{https://ui.adsabs.harvard.edu/abs/2016ApJ...818...40M}{\color{blue}818},
  40

\bibitem[{{Martin} {et~al.}(2022){Martin}, {Venn}, {Aguado}, {Starkenburg},
  {Gonz{\'a}lez Hern{\'a}ndez}, {Ibata}, {Bonifacio}, {Caffau}, {Sestito},
  {Arentsen}, {Allende Prieto}, {Carlberg}, {Fabbro}, {Fouesneau}, {Hill},
  {Jablonka}, {Kordopatis}, {Lardo}, {Malhan}, {Mashonkina}, {McConnachie},
  {Navarro}, {S{\'a}nchez-Janssen}, {Thomas}, {Yuan}, \&
  {Mucciarelli}}]{Martin2022_C19}
{Martin}, N.~F., {Venn}, K.~A., {Aguado}, D.~S., {et~al.} 2022,
  \href{http://dx.doi.org/10.1038/s41586-021-04162-2}{\color{magenta}\nat},
  \href{https://ui.adsabs.harvard.edu/abs/2022Natur.601...45M}{\color{blue}601},
  45

\bibitem[{{Mateo}(1998)}]{Mateo1998}
{Mateo}, M.~L. 1998,
  \href{http://dx.doi.org/10.1146/annurev.astro.36.1.435}{\color{magenta}\araa},
  \href{https://ui.adsabs.harvard.edu/abs/1998ARA&A..36..435M}{\color{blue}36},
  435

\bibitem[{{Mau} {et~al.}(2020){Mau}, {Cerny}, {Pace}, {Choi}, {Drlica-Wagner},
  {Santana-Silva}, {Riley}, {Erkal}, {Stringfellow}, {Adam{\'o}w}, {Carlin},
  {Gruendl}, {Hernandez-Lang}, {Kuropatkin}, {Li}, {Mart{\'\i}nez-V{\'a}zquez},
  {Morganson}, {Mutlu-Pakdil}, {Neilsen}, {Nidever}, {Olsen}, {Sand},
  {Tollerud}, {Tucker}, {Yanny}, {Zenteno}, {Allam}, {Barkhouse}, {Bechtol},
  {Bell}, {Balaji}, {Crnojevi{\'c}}, {Esteves}, {Ferguson}, {Gallart},
  {Hughes}, {James}, {Jethwa}, {Johnson}, {Kuehn}, {Majewski}, {Mao},
  {Massana}, {McNanna}, {Monachesi}, {Nadler}, {No{\"e}l}, {Palmese},
  {Paz-Chinchon}, {Pieres}, {Sanchez}, {Shipp}, {Simon}, {Soares-Santos},
  {Tavangar}, {van der Marel}, {Vivas}, {Walker}, \& {Wechsler}}]{Mau2020}
{Mau}, S., {Cerny}, W., {Pace}, A.~B., {et~al.} 2020,
  \href{http://dx.doi.org/10.3847/1538-4357/ab6c67}{\color{magenta}\apj},
  \href{https://ui.adsabs.harvard.edu/abs/2020ApJ...890..136M}{\color{blue}890},
  136

\bibitem[{{McConnachie}(2012)}]{McConnachie2012}
{McConnachie}, A.~W. 2012,
  \href{http://dx.doi.org/10.1088/0004-6256/144/1/4}{\color{magenta}\aj},
  \href{http://adsabs.harvard.edu/abs/2012AJ....144....4M}{\color{blue}144}, 4

\bibitem[{{McConnachie} \& {C{\^o}t{\'e}}(2010)}]{McConnachieCote2010}
{McConnachie}, A.~W., \& {C{\^o}t{\'e}}, P. 2010,
  \href{http://dx.doi.org/10.1088/2041-8205/722/2/L209}{\color{magenta}\apjl},
  \href{http://adsabs.harvard.edu/abs/2010ApJ...722L.209M}{\color{blue}722},
  L209

\bibitem[{{Namekata} {et~al.}(2018){Namekata}, {Iwasawa}, {Nitadori},
  {Tanikawa}, {Muranushi}, {Wang}, {Hosono}, {Nomura}, \&
  {Makino}}]{Namekata2018}
{Namekata}, D., {Iwasawa}, M., {Nitadori}, K., {et~al.} 2018,
  \href{http://dx.doi.org/10.1093/pasj/psy062}{\color{magenta}\pasj},
  \href{https://ui.adsabs.harvard.edu/abs/2018PASJ...70...70N}{\color{blue}70},
  70

\bibitem[{{Navarro} {et~al.}(1996){Navarro}, {Frenk}, \&
  {White}}]{Navarro1996a}
{Navarro}, J.~F., {Frenk}, C.~S., \& {White}, S. D.~M. 1996,
  \href{http://dx.doi.org/10.1086/177173}{\color{magenta}\apj},
  \href{https://ui.adsabs.harvard.edu/abs/1996ApJ...462..563N}{\color{blue}462},
  563

\bibitem[{{Navarro} {et~al.}(1997){Navarro}, {Frenk}, \& {White}}]{Navarro1997}
{Navarro}, J.~F., {Frenk}, C.~S., \& {White}, S.~D.~M. 1997,
  \href{http://dx.doi.org/10.1086/304888}{\color{magenta}ApJ},
  \href{http://adsabs.harvard.edu/abs/1997ApJ...490..493N}{\color{blue}490},
  493

\bibitem[{{Pe{\~n}arrubia}(2019)}]{Penarrubia2019Scattering}
{Pe{\~n}arrubia}, J. 2019,
  \href{http://dx.doi.org/10.1093/mnras/stz2648}{\color{magenta}\mnras},
  \href{https://ui.adsabs.harvard.edu/abs/2019MNRAS.490.1044P}{\color{blue}490},
  1044

\bibitem[{{Pe{\~n}arrubia}(2021)}]{Penarrubia2021}
---. 2021,
  \href{http://dx.doi.org/10.1093/mnras/staa3700}{\color{magenta}\mnras},
  \href{https://ui.adsabs.harvard.edu/abs/2021MNRAS.501.3670P}{\color{blue}501},
  3670

\bibitem[{{Pe{\~n}arrubia}(2023)}]{Penarrubia2023}
---. 2023,
  \href{http://dx.doi.org/10.1093/mnras/stac3642}{\color{magenta}\mnras},
  \href{https://ui.adsabs.harvard.edu/abs/2023MNRAS.519.1955P}{\color{blue}519},
  1955

\bibitem[{{Pe{\~n}arrubia} {et~al.}(2010){Pe{\~n}arrubia}, {Benson}, {Walker},
  {Gilmore}, {McConnachie}, \& {Mayer}}]{Penarrubia2010}
{Pe{\~n}arrubia}, J., {Benson}, A.~J., {Walker}, M.~G., {et~al.} 2010,
  \href{http://dx.doi.org/10.1111/j.1365-2966.2010.16762.x}{\color{magenta}MNRAS},
  \href{http://adsabs.harvard.edu/abs/2010MNRAS.406.1290P}{\color{blue}406},
  1290

\bibitem[{{Pe{\~n}arrubia} {et~al.}(2025){Pe{\~n}arrubia}, {Errani}, {Vitral},
  \& {Walker}}]{Penarrubia2025}
{Pe{\~n}arrubia}, J., {Errani}, R., {Vitral}, E., \& {Walker}, M.~G. 2025,
  \href{http://dx.doi.org/10.48550/arXiv.2506.03904}{\color{magenta}arXiv:2506.03904}

\bibitem[{{Pe{\~n}arrubia} {et~al.}(2024){Pe{\~n}arrubia}, {Errani}, {Walker},
  {Gieles}, \& {Boekholt}}]{Penarrubia2024}
{Pe{\~n}arrubia}, J., {Errani}, R., {Walker}, M.~G., {Gieles}, M., \&
  {Boekholt}, T. C.~N. 2024,
  \href{http://dx.doi.org/10.1093/mnras/stae1961}{\color{magenta}\mnras},
  \href{https://ui.adsabs.harvard.edu/abs/2024MNRAS.533.3263P}{\color{blue}533},
  3263

\bibitem[{{Pe{\~n}arrubia} {et~al.}(2016){Pe{\~n}arrubia}, {Ludlow},
  {Chanam{\'e}}, \& {Walker}}]{PenarrubiaLudow2016}
{Pe{\~n}arrubia}, J., {Ludlow}, A.~D., {Chanam{\'e}}, J., \& {Walker}, M.~G.
  2016, \href{http://dx.doi.org/10.1093/mnrasl/slw090}{\color{magenta}\mnras},
  \href{http://adsabs.harvard.edu/abs/2016MNRAS.461L..72P}{\color{blue}461},
  L72

\bibitem[{{Petit} \& {Henon}(1986)}]{PetitHenon1986}
{Petit}, J.~M., \& {Henon}, M. 1986,
  \href{http://dx.doi.org/10.1016/0019-1035(86)90089-8}{\color{magenta}\icarus},
  \href{https://ui.adsabs.harvard.edu/abs/1986Icar...66..536P}{\color{blue}66},
  536

\bibitem[{{Safarzadeh} {et~al.}(2022){Safarzadeh}, {Simon}, \&
  {Loeb}}]{Safarzadeh2022}
{Safarzadeh}, M., {Simon}, J.~D., \& {Loeb}, A. 2022,
  \href{http://dx.doi.org/10.3847/1538-4357/ac626e}{\color{magenta}\apj},
  \href{https://ui.adsabs.harvard.edu/abs/2022ApJ...930...54S}{\color{blue}930},
  54

\bibitem[{{Safarzadeh} \& {Spergel}(2020)}]{Safarzadeh2020}
{Safarzadeh}, M., \& {Spergel}, D.~N. 2020,
  \href{http://dx.doi.org/10.3847/1538-4357/ab7db2}{\color{magenta}\apj},
  \href{https://ui.adsabs.harvard.edu/abs/2020ApJ...893...21S}{\color{blue}893},
  21

\bibitem[{{Shariat} {et~al.}(2025){Shariat}, {El-Badry}, {Gennaro}, {Ding},
  {Simon}, {Avila}, {Calamida}, {Cassisi}, {Correnti}, {Weisz}, {Geha},
  {Kirby}, {Brown}, {Ricotti}, {McQuinn}, {Kallivayalil}, {Gilbert},
  {Pacifici}, {Guhathakurta}, {Crnojevi{\'c}}, {Boyer}, {Beaton}, {Chandra},
  {Cohen}, {Renzini}, {Savino}, \& {Tollerud}}]{Cheyanne2025}
{Shariat}, C., {El-Badry}, K., {Gennaro}, M., {et~al.} 2025,
  \href{http://dx.doi.org/10.48550/arXiv.2509.04555}{\color{magenta}arXiv:2509.04555}

\bibitem[{{Simon} {et~al.}(2024){Simon}, {Li}, {Ji}, {Pace}, {Hansen}, {Cerny},
  {Escala}, {Koposov}, {Drlica-Wagner}, {Mau}, \& {Kirby}}]{Simon2024}
{Simon}, J.~D., {Li}, T.~S., {Ji}, A.~P., {et~al.} 2024,
  \href{http://dx.doi.org/10.3847/1538-4357/ad85dd}{\color{magenta}\apj},
  \href{https://ui.adsabs.harvard.edu/abs/2024ApJ...976..256S}{\color{blue}976},
  256

\bibitem[{{Smith} {et~al.}(2024){Smith}, {Cerny}, {Hayes}, {Sestito}, {Jensen},
  {McConnachie}, {Geha}, {Navarro}, {Li}, {Cuillandre}, {Errani}, {Chambers},
  {Gwyn}, {Hammer}, {Hudson}, {Magnier}, \& {Martin}}]{Smith2024}
{Smith}, S. E.~T., {Cerny}, W., {Hayes}, C.~R., {et~al.} 2024,
  \href{http://dx.doi.org/10.3847/1538-4357/ad0d9f}{\color{magenta}\apj},
  \href{https://ui.adsabs.harvard.edu/abs/2024ApJ...961...92S}{\color{blue}961},
  92

\bibitem[{{Spencer} {et~al.}(2017){Spencer}, {Mateo}, {Walker}, {Olszewski},
  {McConnachie}, {Kirby}, \& {Koch}}]{Spencer2018}
{Spencer}, M.~E., {Mateo}, M., {Walker}, M.~G., {et~al.} 2017,
  \href{http://dx.doi.org/10.3847/1538-3881/aa6d51}{\color{magenta}\aj},
  \href{https://ui.adsabs.harvard.edu/abs/2017AJ....153..254S}{\color{blue}153},
  254

\bibitem[{{Spitzer}(1987)}]{Spitzer1987book}
{Spitzer}, L. 1987, {Dynamical evolution of globular clusters} (Princeton
  University Press)

\bibitem[{{Spitzer} \& {Thuan}(1972)}]{SpitzerThuan1972}
{Spitzer}, Jr., L., \& {Thuan}, T.~X. 1972,
  \href{http://dx.doi.org/10.1086/151537}{\color{magenta}\apj},
  \href{https://ui.adsabs.harvard.edu/abs/1972ApJ...175...31S}{\color{blue}175},
  31

\bibitem[{{Springel}(2005)}]{Springel2005Gadget}
{Springel}, V. 2005,
  \href{http://dx.doi.org/10.1111/j.1365-2966.2005.09655.x}{\color{magenta}\mnras},
  \href{https://ui.adsabs.harvard.edu/abs/2005MNRAS.364.1105S}{\color{blue}364},
  1105

\bibitem[{{St{\"u}cker} {et~al.}(2023){St{\"u}cker}, {Ogiya}, {Angulo},
  {Aguirre-Santaella}, \& {S{\'a}nchez-Conde}}]{Stuecker2023}
{St{\"u}cker}, J., {Ogiya}, G., {Angulo}, R.~E., {Aguirre-Santaella}, A., \&
  {S{\'a}nchez-Conde}, M.~A. 2023,
  \href{http://dx.doi.org/10.1093/mnras/stad844}{\color{magenta}\mnras},
  \href{https://ui.adsabs.harvard.edu/abs/2023MNRAS.521.4432S}{\color{blue}521},
  4432

\bibitem[{{van den Bosch} {et~al.}(2018){van den Bosch}, {Ogiya}, {Hahn}, \&
  {Burkert}}]{vdb2018}
{van den Bosch}, F.~C., {Ogiya}, G., {Hahn}, O., \& {Burkert}, A. 2018,
  \href{http://dx.doi.org/10.1093/mnras/stx2956}{\color{magenta}\mnras},
  \href{http://adsabs.harvard.edu/abs/2018MNRAS.474.3043V}{\color{blue}474},
  3043

\bibitem[{{Venn} {et~al.}(2004){Venn}, {Irwin}, {Shetrone}, {Tout}, {Hill}, \&
  {Tolstoy}}]{Venn2004}
{Venn}, K.~A., {Irwin}, M., {Shetrone}, M.~D., {et~al.} 2004,
  \href{http://dx.doi.org/10.1086/422734}{\color{magenta}\aj},
  \href{https://ui.adsabs.harvard.edu/abs/2004AJ....128.1177V}{\color{blue}128},
  1177

\bibitem[{{Walker} {et~al.}(2007){Walker}, {Mateo}, {Olszewski}, {Gnedin},
  {Wang}, {Sen}, \& {Woodroofe}}]{Walker2007}
{Walker}, M.~G., {Mateo}, M., {Olszewski}, E.~W., {et~al.} 2007,
  \href{http://dx.doi.org/10.1086/521998}{\color{magenta}\apjl},
  \href{http://adsabs.harvard.edu/abs/2007ApJ...667L..53W}{\color{blue}667},
  L53

\bibitem[{{Wang} {et~al.}(2020{\natexlab{a}}){Wang}, {Iwasawa}, {Nitadori}, \&
  {Makino}}]{Wang2020_PETAR}
{Wang}, L., {Iwasawa}, M., {Nitadori}, K., \& {Makino}, J. 2020{\natexlab{a}},
  \href{http://dx.doi.org/10.1093/mnras/staa1915}{\color{magenta}\mnras},
  \href{https://ui.adsabs.harvard.edu/abs/2020MNRAS.497..536W}{\color{blue}497},
  536

\bibitem[{{Wang} {et~al.}(2020{\natexlab{b}}){Wang}, {Nitadori}, \&
  {Makino}}]{Wang2020_SDAR}
{Wang}, L., {Nitadori}, K., \& {Makino}, J. 2020{\natexlab{b}},
  \href{http://dx.doi.org/10.1093/mnras/staa480}{\color{magenta}\mnras},
  \href{https://ui.adsabs.harvard.edu/abs/2020MNRAS.493.3398W}{\color{blue}493},
  3398

\bibitem[{{White} \& {Rees}(1978)}]{WhiteRees1978}
{White}, S.~D.~M., \& {Rees}, M.~J. 1978,
  \href{http://dx.doi.org/10.1093/mnras/183.3.341}{\color{magenta}\mnras},
  \href{http://adsabs.harvard.edu/abs/1978MNRAS.183..341W}{\color{blue}183},
  341

\bibitem[{{Yuan} {et~al.}(2022){Yuan}, {Martin}, {Ibata}, {Caffau},
  {Bonifacio}, {Mashonkina}, {Errani}, {Doliva-Dolinsky}, {Starkenburg},
  {Venn}, {Arentsen}, {Aguado}, {Bellazzini}, {Famaey}, {Fouesneau},
  {Gonz{\'a}lez Hern{\'a}ndez}, {Jablonka}, {Lardo}, {Malhan}, {Navarro},
  {S{\'a}nchez Janssen}, {Sestito}, {Thomas}, {Viswanathan}, \&
  {Vitali}}]{Yuan2022}
{Yuan}, Z., {Martin}, N.~F., {Ibata}, R.~A., {et~al.} 2022,
  \href{http://dx.doi.org/10.1093/mnras/stac1399}{\color{magenta}\mnras},
  \href{https://ui.adsabs.harvard.edu/abs/2022MNRAS.tmp.1369Y}{\color{blue}514},
  1664

\bibitem[{{Zaremba} {et~al.}(2025){Zaremba}, {Venn}, {Hayes}, {Errani},
  {Cornejo}, {Glover}, {Jensen}, {McConnachie}, {Navarro}, {Pazder}, {Sestito},
  {Anthony}, {Andersen}, {Baker}, {Chin}, {Churilov}, {Diaz}, {Farrell},
  {Firpo}, {Gomez-Jimenez}, {Henderson}, {Kalari}, {Lawrence}, {Margheim},
  {Miller}, {Robertson}, {Ruiz-Carmona}, {Silversides}, {Silva}, {Young}, \&
  {Zhelem}}]{Zaremba2025}
{Zaremba}, D., {Venn}, K., {Hayes}, C.~R., {et~al.} 2025,
  \href{http://dx.doi.org/10.3847/1538-4357/add5f9}{\color{magenta}\apj},
  \href{https://ui.adsabs.harvard.edu/abs/2025ApJ...987..217Z}{\color{blue}987},
  217

\bibitem[{{Zhang} \& {Amaro Seoane}(2025)}]{ZhangAmaro2025}
{Zhang}, F., \& {Amaro Seoane}, P. 2025,
  \href{http://dx.doi.org/10.3847/1538-4357/adaa7a}{\color{magenta}\apj},
  \href{https://ui.adsabs.harvard.edu/abs/2025ApJ...980..210Z}{\color{blue}980},
  210

\end{thebibliography}

\label{lastpage}
\end{document}